\documentclass[preprintnumbers,aps,prb,twocolumn,showkeys]{revtex4}
\usepackage{tabularx}
\usepackage{bm}
\usepackage{dcolumn} 
\usepackage{euscript}
\usepackage{epsfig,psfrag,subfigure,subeqnarray}
\usepackage{graphicx}
\usepackage{color}
\usepackage{amsfonts}
\usepackage{amsmath}
\usepackage{amsbsy}

\def\lan{\langle}
\def\ran{\rangle}
\def\llan{\langle\langle}
\def\rran{\rangle\rangle}

\def\va{\varepsilon}
\def\sa{\sigma}

\def\bsa{{\bar{\sigma}}}

\def\sp{{ \sa^{\prime }}}

\def\bk {{\bf k}}
\def\bp {{\bf p}}

\newcommand{\bd}{\begin{equation}}
\newcommand{\ed}{\end{equation}}
\newcommand{\be}{\begin{equation}}
\newcommand{\ee}{\end{equation}}
\newcommand{\bt}{\begin{split}}
\newcommand{\et}{\end{split}}

\newcommand{\bn}{\begin{align}}
\newcommand{\en}{\end{align}}

\newcommand{\bea}{\begin{eqnarray}}
\newcommand{\eea}{\end{eqnarray}}
\newcommand{\ba}{\begin{array}}
\newcommand{\ea}{\end{array}}
\newcommand{\nn}{\nonumber}

\newcommand{\bra}{\langle\langle }
\newcommand{\ket}{\rangle\rangle }

\begin{document}
\title{Bistability in the Tunnelling Current through a Ring of $N$ Coupled Quantum Dots}

\author{Shiue-Yuan Shiau and Yia-Chung Chang} 
\email{yiachang@gate.sinica.edu.tw}

\affiliation{Research Center for Applied Sciences, Academia Sinica, Taipei, 115 Taiwan}
\author{David M.-T. Kuo}
\affiliation{Department of Electrical Engineering and Department of Physics, National Central University, Chungli, 320 Taiwan}
\date{\today}

\begin{abstract}
We study bistability in the electron transport through a ring of $N$ coupled quantum dots with two orbitals in each dot. One orbital is localized (called $b$ orbital) and coupling of the $b$ orbitals in any two dots is negligible; the other is delocalized in the plane of the ring (called $d$ orbital), due to coupling of the $d$ orbitals in the neighboring dots, as described by a tight-binding model. The $d$ orbitals thereby form a band with finite width. The $b$ and $d$ orbitals are connected to the source and drain electrodes with a voltage bias $V$, allowing the electron tunnelling. Tunnelling current is calculated by using a nonequilibrium Green function method recently developed to treat nanostructures with multiple energy levels. We find a bistable effect in the tunnelling current as a function of bias $V$, when the size $N\gtrsim 50$; this effect scales with the size $N$ and becomes sizable at $N\sim 100$. The temperature effect on bistability is also discussed. In comparison, mean-field treatment tends to overestimate the bistable effect.
\end{abstract}

\maketitle

\section{Introduction}
Bistability (or hysteresis) has always stood as one of the most fascinating phenomena in condensed matter physics: its original mechanism can be either magnetic as in novel materials like metal-ion cluster\cite{Sessoli1993} and organic radical crystals\cite{Wataru1999}, or electric as in resonant tunnelling diodes\cite{Goldman1987}. For resonant tunnelling structures, their intrinsic bistability behaviour has continued to attract attention\cite{Zhao2000}. Recently, a novel optical bistability in polariton diodes was also reported\cite{Bajoni2008}. Besides being physically intriguing in its own right, bistable characteristic can be used to fabricate memory devices\cite{Ouyang2004}. Nanoscale devices exhibiting such property can therefore have potential application for low-cost, low-power and high-density memory storage. \

Recent theoretical studies have explored the possibility of making a memory device of a highly degenerate molecular quantum dot\cite{AB2003} or of a single-level molecular junction\cite{GRN2005,MD}; both cases rely critically on strong electron-phonon interaction (polaron effect) in order to make possible effective attractive electron-electron interaction. However, two experiments measuring transport through a carbon nanotube quantum dot\cite{Moon2007} with an eightfold degenerate state, and through a single spherical PbSe quantum dot\cite{Liljeroth2006} with a sixfold degenerate state did not show any bistable characteristics. These results indicate that electron-phonon interaction may not be strong enough in carbon nanotube quantum dots or PbSe quantum dots to induce attractive electron-electron interaction. It remains inconclusive whether a single quantum dot junction can exhibit bistability.\

Recent experiments have demonstrated that semiconductor\cite{Murray1995} and metallic\cite{Collier1997} quantum dots can be chemically engineered into geometrically-ordered superlattices (nanocrystals). Nanoscale manipulation allows to control the lattice constant and quantum dot size with a precision limited by atomic roughness\cite{Murray1995}, to make a quantum dot array (QDA) evolve from a Coulomb blockade regime to a semiconducting regime by tuning the interdot coupling\cite{Romero2005}, and to form a band structure from a QDA\cite{Lobo2009}. As a result, QDAs provide a well-controlled system for study of strong correlation\cite{Mentzel2008} as well as a promising integrated electronic device.\

Previous mean-field study\cite{KuoChang09} on QDA junctions described by the Falicov-Kim model\cite{FK1969,FK1970} predicted bistability behaviour in current-voltage characteristics, which makes devices made of QDAs a good candidate for low-power high-density memory devices. Each dot in the QDA junction contains localized and delocalized states with on-site repulsive Coulomb interaction. Nonequilibrium mean-field theory\cite{KuoChang09} produced two coupled transcendental equations that can be solved for the average single-particle occupation numbers as a function of voltage bias. Meta-stable states are then made possible by the charge transfer between these two kinds of states, leading to bistable tunnelling current. However, in addition to doubt that could be raised against mean-field treatment in strong Coulomb interaction\cite{Horvath2008}, this simple theory can not give any answer to at what size of the QDA junction bistability starts to emerge and how it scales with the system size---an important issue for experimentalists who intend to realize bistability in such system and possibly optimize it for future manufacturing. \

In this paper, to study the size effect, we consider a ring of $N$ coupled quantum dots simulated by a multi-level Anderson model. The model is tackled by using a nonequilibrium Green function method\cite{KuoChang07,Chang08} that is able to take into account strong correlations in intra-orbital and inter-orbital states. This method was originally developed to study a nanostructure tunnel junction with multiple energy levels {in the Coulomb blockade regime}. \

Our results show qualitative agreement with those obtained from nonequilibrium mean-field theory: Bistability depends critically on the strength of the Coulomb energy $U$, the tight-binding hopping parameter $t_d$, and the ratio of the left and right tunnelling rates; the Coulomb energy $U$ needs to be large enough for bistability to appear.  However, our method shows that mean-field theory tends to overestimate the bistable effect by incorrectly incorporating higher-energy spectral density of the delocalized states, even though the bias is too low to allow population at high energy. In particular, we show that bistability in the tunnelling current occurs when $N\gtrsim 50$ and becomes sizable around $N\sim 100$ in a QDA system with typical physical parameters; the best $t_d$ value for pronounced bistability is found around $U/4$. This size estimation serves as an important guide for fabrication of QDA junctions aimed at memory devices. Finally, we show that increasing temperature tends to destroy bistability, as does the interdot Coulomb interaction between localized states. To preserve bistability, it is then important to keep the localized states well separated, and for the temperature to be much smaller than the effective tunnelling rates through the localized states.

The paper is organized as follows: Section~\ref{sec:Model} introduces the multi-level Anderson model and the approximations involved. In Sec.\ref{sec:Current} we formulate the tunnelling current and derive its simple linear relation with the average occupation number. In Sec.\ref{sec:GFs}, we discuss the nonequilibrium Green function method. In Sec.\ref{sec:Results}, we present numerical results. We conclude in the end. \

\section{ Multi-level Anderson Model\label{sec:Model}}
\begin{figure}[t]
\centering
\includegraphics[scale=0.22]{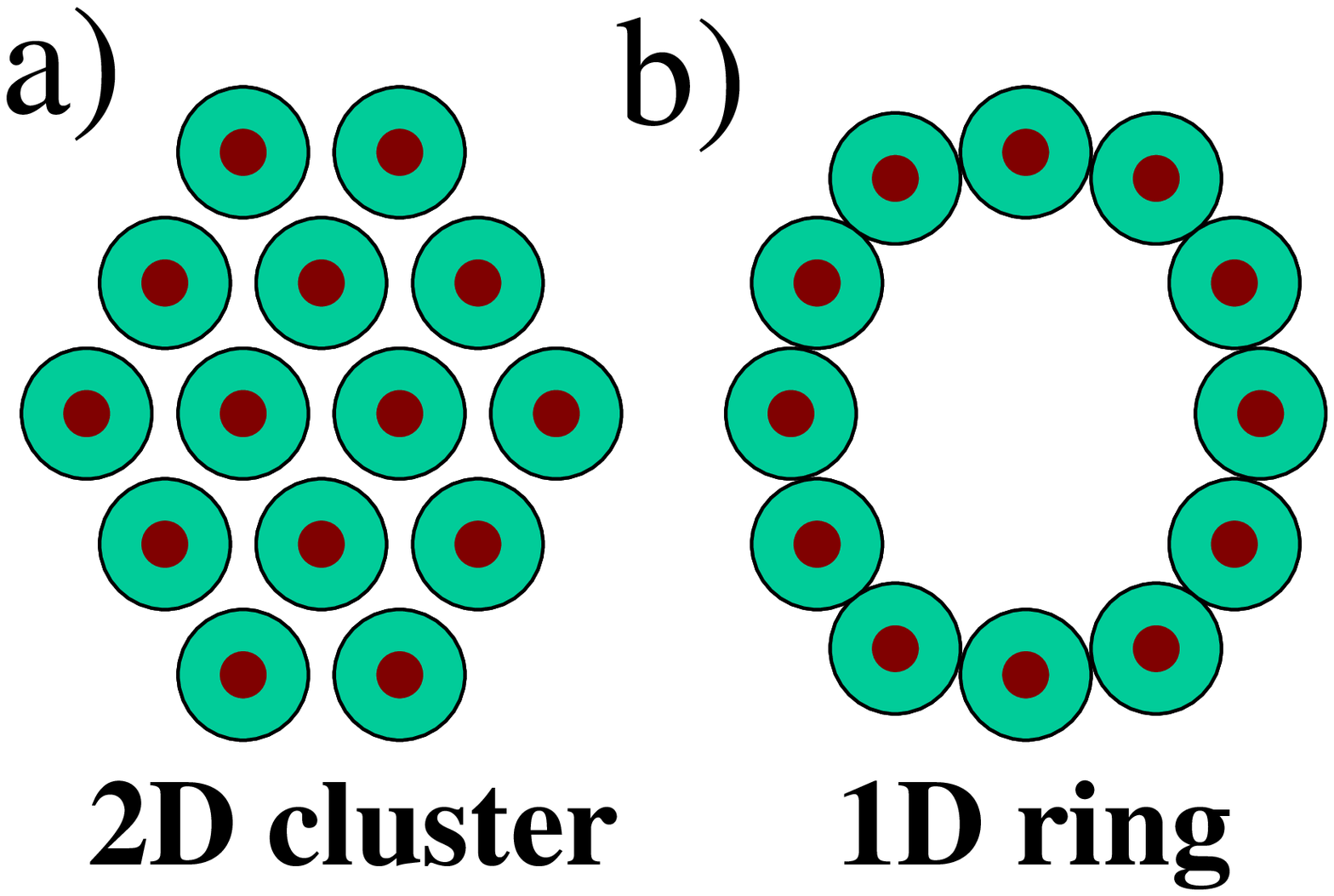}\\
\vspace{-4cm}
\includegraphics[scale=0.3]{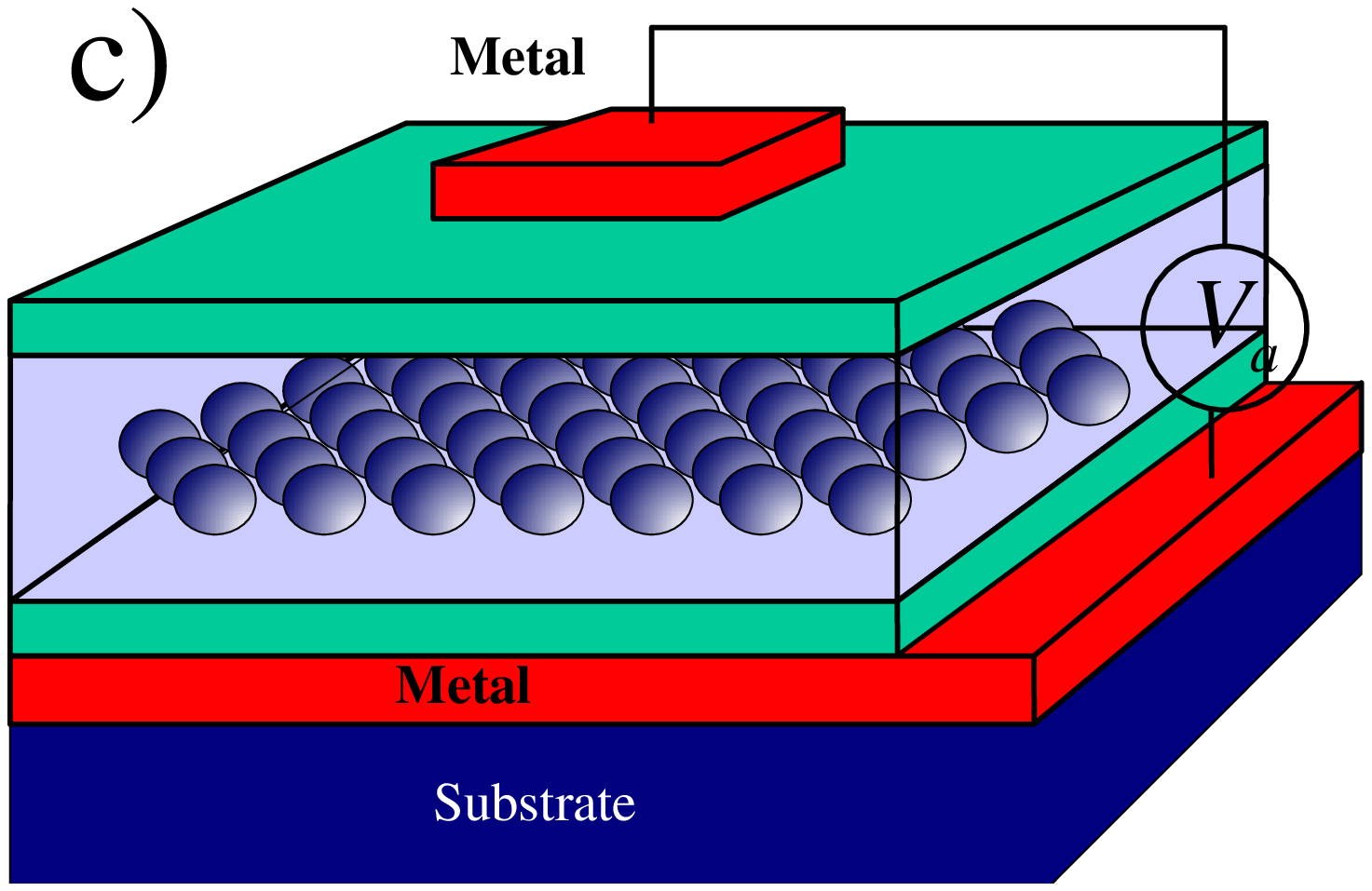} \\
\vspace{-5cm}
\includegraphics[scale=0.17]{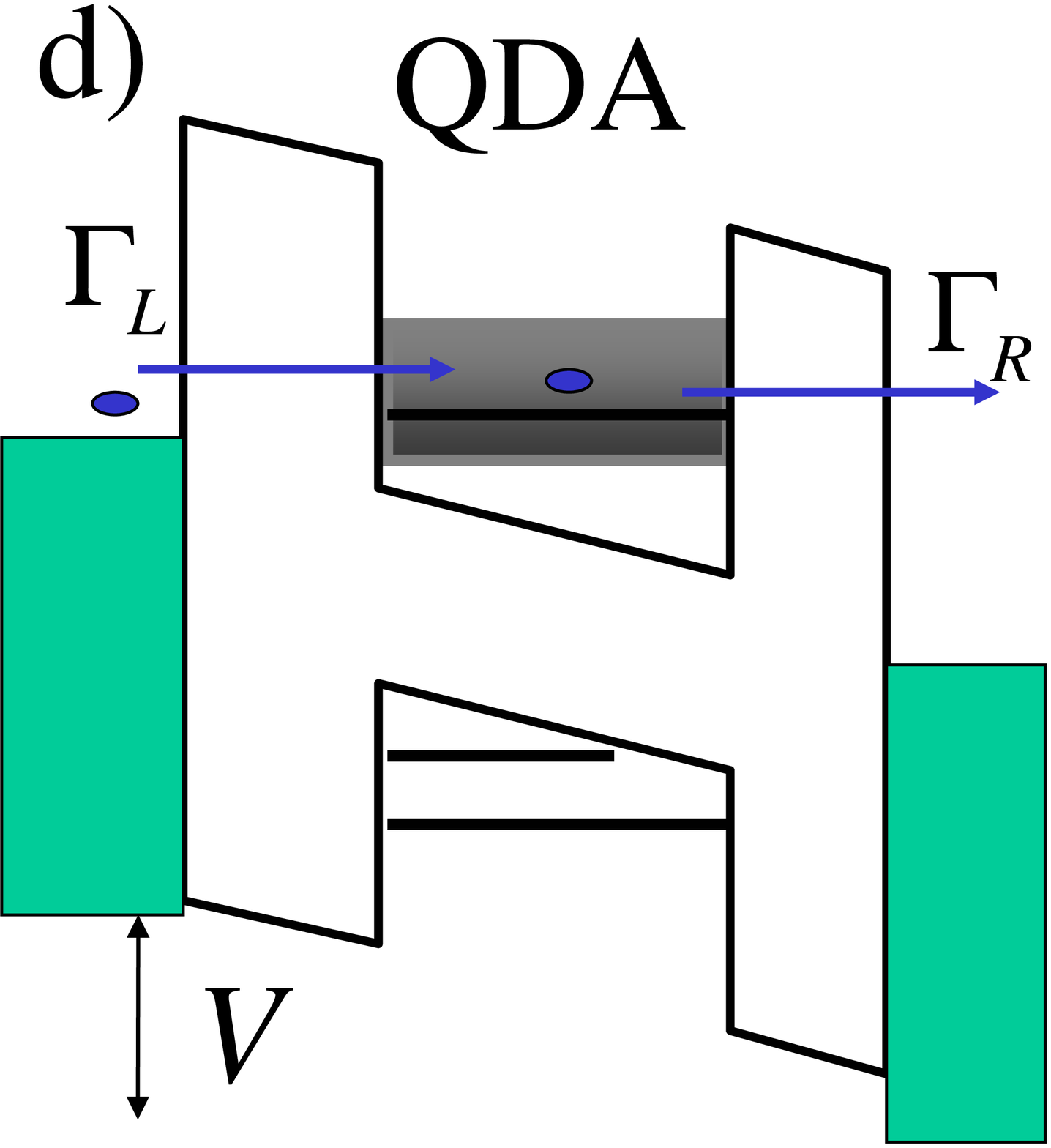} \\
\caption{\protect\small{(Color online)Finite-sized (a) 2D and (b) 1D QDAs. Each dot is a combination of a large dot (green) and a small dot (brown). (c) QDA embedded into an insulated matrix sandwiched by two metallic leads. (d) Schematic of electrons tunnelling through a QDA in the presence of an external bias $V$. The QDA consists of localized $b$ states (discrete black lines) and delocalized $d$ states forming a finite band (grey area).
  }}\label{fig:Conf_QDA}
\end{figure}

To observe bistability in the tunnelling current through a QDA, an experimental setup\cite{KuoChang09} was proposed,  see Fig.~\ref{fig:Conf_QDA}(c). The QDA is embedded in an insulated layer sandwiched by two metallic leads. To these two leads a voltage bias is applied and current is measured between them. The system can be either 2D or 1D QDA, as shown in Fig.~\ref{fig:Conf_QDA}(a) and Fig.~\ref{fig:Conf_QDA}(b) for finite-sized 2D and 1D QDAs, respectively. Possible QDA candidates are molecular solids and nanocrystals that potentially can display bulk properties (e.g. band formation).  \

To study the size effect, we consider a 1D ring-like array of $N$ coupled quantum dots (see Fig.~\ref{fig:Conf_QDA}(b)). Each QD is in fact a combination of a small dot (which hosts a localized orbital, called $b$ orbital) and a larger dot (which hosts a delocalized orbital, called $d$ orbital), e.g. in a core-shell like structure. We then consider the case that the $b$ orbital is either $s$-like or $p_z$-like, and the $d$ orbital is $p_x$-like or $p_y$-like, so they have different parities in the plane of QDA. The $d$ orbitals of two neighboring dots are coupled via a tight-binding interaction $t_{nd}$. The $d$ and $b$ orbitals are connected to two metallic leads (labeled by $\alpha= L,~R$) with an applied voltage (see Fig.~\ref{fig:Conf_QDA}(c)). Both the intradot and interdot Coulomb interactions $H_U$ and $H_{xU}$ are taken into account. We therefore use a generalized multi-level Anderson model\cite{Anderson1961} to describe such system:
\bea
H&=&H_{leads}+H_T+H_{QDA}+H_U+H_{xU},\label{eq:Hamil01}\\
H_{leads}&=&\sum_{\alpha \bk \sa}\va_{\bk } c^\dag_{\alpha \bk \sa}c_{\alpha \bk \sa},\label{eq:Hamil_lead}\\
H_T&=& \frac{1}{\sqrt{N}}\sum_{\alpha \bk \sa}\sum_{n=1}^{N}t_{\alpha \bk nd\sa} d^\dag_{n \sa} c_{\alpha \bk\sa}\nn\\&&+\frac{1}{\sqrt{N}}\sum_{\alpha \bk \sa}\sum_{n=1}^{N} t_{\alpha \bk n b\sa}b^\dag_{n \sa} c_{\alpha \bk\sa}+h.c.,\label{eq:Hamil02}\\
H_{QDA}&=&\sum_{n=1}^{N}\sum_\sa\va_{nb\sa}b^\dag_{n\sa}b_{n\sa}+\sum_{n=1}^{N}\sum_\sa\va_{nd\sa}d^\dag_{n\sa}d_{n\sa}\nn\\&&+\sum_{n=1}^{N} \sum_\sa t_{nd} d^\dag_{n\sa} d_{n+1 \sa}\label{eq:Hamil03},\\
H_U&=& \sum_{n=1}^{N} U_{ndd}n_{n d\uparrow}n_{n d\downarrow}+\sum_{n=1}^{N} U_{nbb}n_{ n b\uparrow}n_{ nb\downarrow}\nn\\&&+\sum_{n=1}^{N}U_{ndb} \sum_{\sa,\sa^\prime} n_{n b\sa} n_{ nd\sa^\prime},\label{eq:Hamil04}\\
H_{xU}&=&\frac{1}{2}\sum_{\sa,\sp}\sum_{n,m=1\atop n\not=m}^{N} {\mathcal V}_{bb}(R_{nm})n_{n b\sa}n_{m b\sp}\nn\\
&&+\sum_{\sa,\sp}\sum_{n,m=1\atop n\not=m}^{N} {\mathcal V}_{db}(R_{nm})  n_{n b\sa} n_{ md\sa^\prime}\nn\\&&+ \frac{1}{2}\sum_{\sa,\sp}\sum_{n,m=1\atop n\not=m}^{N} {\mathcal V}_{dd}(R_{nm})n_{n d\sa}n_{m d\sp},\label{eq:Hamil05}
\eea

where $c_{\alpha \bk\sa }^{\dag }(c_{\alpha \bk\sa })$ is the creation (annihilation) operator of a free electron of momentum $\bk$ and spin $\sigma(=\pm 1)$ in the $\alpha$-lead with energy $\va_\bk$; $b_{n\sigma }^{\dag}(b_{n\sigma })$ and $d_{n\sigma }^{\dag}(d_{n\sigma })$ are the creation (annihilation) operators of electrons of spin $\sigma$ in the $b$ and $d$ orbitals of the $n$th quantum dot with energies $\varepsilon _{nb\sa}$ and $\varepsilon _{nd\sa}$, respectively. Their respective number operators are $n_{nb\sa}=b^\dag_{n\sa}b_{n\sa}$ and $n_{nd\sa}=d^\dag_{n\sa}d_{n\sa}$. $U_{njl} (j,l=d,b)$ denotes the intradot Coulomb interaction between two electrons of $j$ and $l$ orbitals in the $n$th dot, while ${\mathcal V}_{jl}(R_{nm})$ the interdot Coulomb interaction between the $n$th and the $m$th dots. {For the latter, we shall adopt a screened Coulomb potential, i.e., Yukawa-type potential, which reads
\[{\mathcal V}_{jl}(R_{nm})=U_{xjl}\frac{R_1}{R_{nm}}{\rm exp}\left[-\frac{R_{nm}}{L_{jl}}\right],\]
where $R_{nm}(=|{{\bf R}_n-{\bf R}_m}|)$ is the distance of the two dots and $R_1$ is the nearest-neighbor distance. $L_{jl}$ is the screening length.} The constant $U_{xjl}$ denotes the Coulomb interaction between nearest neighboring dots, typically a fraction of $U_{dd}$ or $U_{db}$. $t_{nd}$ is the hopping interaction between the $d$ orbitals in two neighboring $n$th and $(n+1)$th dots, and $t_{\alpha \bk nl \sa}$ is the tunnelling matrix element between the state $|\bk \sa\rangle$ in the $\alpha$ lead and the state $|nl\sa\rangle$ in the $n$th dot. \

In order to render the Hamiltonian (\ref{eq:Hamil01}) more tractable, we make some general assumptions: We assume the electron wave function in each dot takes the same form, thus allowing to be site-independent orbital energy levels $\va_{nl\sa}=\va_{l\sa}$, and Coulomb energies $U_{njl} =U_{jl}(j,l=d,b)$. We further consider the small dots to lie far apart from each other so that due to the large distance as well as the screening from electrons in the larger dots {and background charges---which results in a small screening length---}the interdot Coulomb interaction for the localized states is small compared with their intradot Coulomb interaction. For simplicity, we shall consider only the Coulomb interaction between two neighboring $b$ orbitals, i.e., {${\mathcal V}_{bb}(R_{mn})=\delta_{m,n\pm 1}\widetilde U_{xbb}$ where $\widetilde U_{xbb}=U_{xbb}{\rm exp}[-R_1/L_{bb}]$}.  We then take the tight-binding coupling to be real and site independent; so $t_{nd}=t_d$.\

In real space, $|nd\sa\ran$ states are coupled via a tight-binding interaction (\ref{eq:Hamil03}), which we consider to be strong in favor of formation of a band with finite width\cite{Lobo2009}. This motivated us to first transform the $|nd\sa\ran$ states into band states, which are eigenstates of the rotation operators that leave the ring of dots invariant. To this end, we define $d$-state operators in momentum space
\bea
d_{n\sa}&=&\frac{1}{\sqrt{N}}\sum_{p=1}^N e^{i{p \phi_n}}d_{p\sa},\label{eq:basis_tran01}\\
d_{p\sa}&=&\frac{1}{\sqrt{N}}\sum_{n=1}^{N} e^{-i{p \phi_n}}d_{n\sa},\label{eq:basis_tran02}
\eea
where $\phi_n=2\pi n/N$. We can then rewrite parts of the Hamiltonian as

\be
H_T=\sum_{\alpha n\bk \sa}t_{\alpha  \bk nb\sa} b^\dag_{n \sa} c_{\alpha \bk\sa}+\sum_{\alpha p \eta \sa}t_{\alpha  d\sa} d^\dag_{p\sa}c_{\alpha p\eta\sa}+h.c.,\label{eq:tran_H01}
\ee
\be
H_{QDA}=\sum_{n=1}^{N}\sum_\sa\va_{b\sa}b^\dag_{n\sa}b_{n\sa}+\sum_{p=1}^N\sum_\sa\va_{dp\sa}d^\dag_{p\sa}d_{p\sa}\label{eq:tran_H02},
\ee
\bea
 H_U&=&U_{bb} \sum_{n=1}^{N}n_{n b\uparrow}n_{nb\downarrow}+\frac{U_{dd}}{N}\sum_{q=1}^N\rho_{d\uparrow}(q)\rho_{d\downarrow}(-q)\nn\\
&&+\frac{U_{db}}{N}\sum_{n=1}^{N}\sum_{\sa,\sp} n_{nb\sa}\sum_{q=1}^Ne^{-i{q\phi_n}} \rho_{d\sp}(q),\label{eq:tran_H03}
\eea
\bea
H_{xU}&=&\widetilde U_{xbb}\sum_{n=1}^{N}\sum_{\sa,\sp}n_{n b\sa}n_{n+1 b\sp}\nn\\
&&+\frac{1}{2N}\sum_{q=1}^N\sum_{\sa,\sp} v_{dd}(q)\rho_{d\sa}(q)\rho_{d\sp}(-q)\nn\\&&+\frac{1}{N}\sum_{n=1}^{N}\sum_{\sa,\sp} n_{nb\sa}\sum_{q=1}^N  v_{db}(q) \rho_{d\sp}(-q),\label{eq:tran_H04}
\eea

where $\rho_{d\sa}(q)=\sum_{p=1}^Nd^\dag_{p+q\sa}d_{p\sa}$ and the $d$-orbital energy levels in momentum space read $\va_{dp\sa}=\va_{d\sa}+2t_d\cos(2\pi p/N)$. We also define
\bea
v_{ij}(q)&=&\sum_{n,n\not=0}e^{-i{q\phi_n}}{\mathcal V}_{ij}({\bf R}_n)\label{eq:Vij}.
\eea
Note that $\sum_q v_{ij}(q)=0$ since in the summation the term with $n=0$ is excluded; this term corresponds to the intra Coulomb interaction. For the lead-dot coupling $t_{\alpha d \sa}$ in $H_T$ (\ref{eq:tran_H01}), after averaging out their site dependence, we have
\bea
&&\frac{1}{\sqrt{N}}\sum_{\alpha \bk \sa}\sum_{n=1}^{N} t_{\alpha \bk nd\sa}d^\dag_{n \sa} c_{\alpha \bk\sa}\nn\\&&=\frac{1}{N}\sum_{\alpha \bk \sa}\sum_{n=1}^{N}t_{\alpha d\sa}e^{i{\bf k_\parallel\cdot R_n}}\sum_\bp e^{-i{\bf p\cdot R_n}}d^\dag_{p\sa}c_{\alpha \bk\sa}\nn\\
&&=  \sum_{\alpha \sa}t_{\alpha d\sa}\sum_{ \bp\bk }\delta{\bf (k_\parallel-p)}d^\dag_{p\sa}c_{\alpha \bk\sa}\nn\\
&&=  \sum_{\alpha p\eta\sa}t_{\alpha d\sa} d^\dag_{p\sa}c_{\alpha p\eta\sa},\label{eq:Hm03}
\eea
where $\bk_\parallel$ denotes the $\bk$ component parallel to the plane of the QDA ring; $\eta$ denotes the additional degeneracy for the conduction electron state $|\bk\ran$. We have absorbed the factor $1/\sqrt{N}$ into $t_{\alpha\bk nb\sa}$. \

To solve the Hamiltonian (\ref{eq:tran_H01},\ref{eq:tran_H02},\ref{eq:tran_H03},\ref{eq:tran_H04}) would be extremely difficult, since the Coulomb interaction between the $d$ orbitals in momentum space is not diagonal. We now assume that the delocalized $d$ states are in a paramagnetic phase; this phase is favored when tight-binding coupling is strong\cite{Mahanbook}, a condition taken throughout our numerical results. In this phase, the delocalized state in each quantum dot has equal probability of occupancy. Thus, on average $\langle\rho_{d\sigma}(q)\rangle=\sum_n \langle n_{nd\sa}\rangle e^{i{q\phi_n}}$ equals zero unless $q=0$. We then approximate $\rho_{d\sigma}(q)$ to be zero if $q\not=0$. \

Let us first look at the interdot Coulomb interaction between the delocalized states. In the paramagnetic phase,
\bea
&&\frac{1}{2N}\sum_{\sa,\sp}\sum_{q=1}^N v_{dd}(q)\rho_{d\sa}(q)\rho_{d\sp}(-q)\nn\nn\\
&&\simeq\frac{v_{dd}(q=0)}{2N}\sum_{\sa,\sp}\rho_{d\sa}(0)\rho_{d\sp}(0).\label{eq:extUdd}
\eea
{We then can use Eq.~(\ref{eq:Vij}) to obtain
\be
v_{dd}(0)\approx U_{xdd}\frac{\pi}{N}\sum_{n=1}^{N-1}\frac{e^{[-NR_1\sin(\pi n/N)/\pi L_{dd}]}}{\sin(\pi n/N)}\label{eq:v_dd}
\ee
for large $N$ in a 1D ring-like QDA. When $L_{dd}\rightarrow \infty$, $v_{dd}(0)\approx 2U_{xdd} (\gamma+{\rm ln}2N/\pi)$, where $\gamma$ denotes Euler's constant.} Hence, in the QDA of large size ($N\sim 100$), we find the total interdot Coulomb interaction to be comparable to the intradot Coulomb interaction ($U_{dd}$) {provided that we neglect the screening effect}. Similar arguments hold for the interdot Coulomb interaction {$v_{db}(0)$} between the localized and the delocalized states. For the interdot Coulomb interaction between localized orbitals, we shall keep only the nearest-neighbor term to examine its effect on the bistability. \

Combining the above results, the Coulomb interactions become
\be
H_U+H_{xU}=U_{bb} \sum_{n=1}^{N}n_{n b\uparrow}n_{nb\downarrow}+\widetilde U_{xbb}\sum_{n=1}^{N}\sum_{\sa,\sp}n_{n b\sa}n_{n+1 b\sp}\nn
\ee
\be
+\frac{\widetilde U_{dd}}{N}\rho_{d\uparrow}(0)\rho_{d\downarrow}(0)+\frac{\widetilde U_{db}}{N}\sum_{n=1}^{N}\sum_{\sa\sp}  n_{nb\sa} \rho_{d\sp}(0),\label{eq:tran_H05}
\ee
where {$\widetilde U_{jl}=U_{jl}+v_{jl}(0)/2$ and $\widetilde U_{xbb}=U_{xbb}{\rm exp}[-R_1/L_{bb}]$}. The Hamiltonian comprising Eqs.~(\ref{eq:Hamil_lead},\ref{eq:tran_H01},\ref{eq:tran_H02},\ref{eq:tran_H05}) will be solved in Sec.\ref{sec:GFs} using a nonequilibrium Green function method\cite{KuoChang07,Chang08}.

\section{Tunnelling Current\label{sec:Current}}
The general expression for the current is a function of the tunneling rate ${\bf\hat \Gamma}$ and the local Green functions ${\bf\hat G^{r,<}}$, which are matrices spanned by the channels (orbital and spin levels)\cite{Meir92,Jau}. To start with, we assume there is no direct coupling between the localized and delocalized states, since their orbital wave functions have different parities. Neglecting possible leaking inter-orbital tunneling current, ${\bf\hat \Gamma}$ and ${\bf\hat G^{r,<}}$ become block diagonal matrices with respect to the localized and delocalized orbitals. The total current can then be divided into two parts, namely, $I=I_{b}+I_{d}$; $I_l$ denotes the current tunnelling through the $l$ orbital (see Fig.~\ref{fig:Conf_QDA} for the schematics of current tunnelling through a QDA.).  \

 We next assume the proportionality of the left and right tunneling
rates for each current, so we can further rewrite the currents through the $b$ and $d$ states in terms of retarded Green functions only. They read\cite{Meir92,Jau,keldysh1964}
\be
I_\ell=\frac{-e}{\hbar}\int_{-W}^W\frac{d\va}{\pi}[f_L(\va)-f_R(\va)] {\rm Im}[tr({\bf\hat \Gamma}\mathcal{\bf\hat G}^r_{\ell,\ell})],\label{eq:I_l}
\ee
where ${\bf\hat \Gamma}={\bf\hat\Gamma}_{L \ell }{\bf\hat\Gamma}_{R\ell}/({\bf\hat\Gamma}_{L \ell }+{\bf\hat\Gamma}_{R\ell})$; $\ell=b,d$. $f_\alpha(\va)=f_F(\va-\mu_\alpha)$ and $\mu_\alpha$ are the Fermi distribution function and the chemical potential of the $\alpha$-lead, respectively. $W$ denotes the half band-width of the two metallic leads. Note that with localized states, we take the trace over the number of dots including spin, while with delocalized states, we take it over the band quantum number $p$($p=1,\cdots, N$).  \

For the localized orbitals, the current contributions from off-diagonal (inter-level) channels are of second order in tunneling rate $\Gamma$, since the off-diagonal Green functions $G^{r}_{n,m}$ for $n\not=m$ ($n, m$ labeling the channels in the QDA) can be expressed in terms of diagonal Green functions multiplied by a self-energy $\Sigma_{n,m}$, that is, $G^{r}_{n,m}\simeq G^{r}_{n,n}\Sigma_{n,m}G^{r}_{m,m}$\cite{Chand2010}. Moreover, neglecting the small real part of the self-energy, the imaginary part $\Gamma_{n,m}$ ({\rm Im}$\Sigma_{n,m}$) reads
\bea
\Gamma_{n,m}&=&\pi\sum_{\alpha\bk} t_{\alpha \bk nb\sa}t^*_{\alpha \bk mb\sa}\delta(\va-\va_\bk)\nn\\
&\simeq&\frac{\Gamma_{ b\sa}(\va)}{2} \frac{\sin(k_FR_{mn})}{k_FR_{mn}},
\eea
where $ t_{\alpha \bk nb\sa}= t_{\alpha b\sa}e^{i{\bf k_\parallel\cdot R_n}}$. $\Gamma_{b\sa}(\va)=\Gamma_{L b\sa}(\va)+\Gamma_{Rb\sa
}(\va)$ with $\Gamma_{\alpha  b\sa}(\va)(=2\pi\sum_{\bk} |t_{\alpha b\sa}|^2\delta(\va-\va_{\bk}))$ being the tunnelling rate from the $|b\sa\ran$ state in the dots to the $\alpha$-lead. Take the Fermi wave vector $k_F\sim 10 nm^{-1}$ for a normal metallic lead with Fermi energy around 5 eV. We see that the factor $ \sin(k_FR_{mn})/k_FR_{mn}$ is small when $R_{mn}\gg 1$nm. Therefore, {in the Coulomb blockade regime where the lead-dot coupling $\Gamma$ is small}, and with dots well separated, we can neglect those small off-diagonal terms in Eq.~(\ref{eq:I_l}) and keep only the diagonal contributions. Assume such to be the case for the localized states. We thus approximate the current $I_b$ to be\
\bea
I_b&\simeq&\frac{-Ne}{\hbar}\sum_\sa\int_{-W}^W\frac{d\va}{\pi}\frac{\Gamma_{L b\sa}(\va)\Gamma_{Rb\sa
}(\va)}{\Gamma_{L b \sa}(\va)+\Gamma_{Rb\sa
}(\va)}\nn\\&& \times[f_L(\va)-f_R(\va)]{\rm Im}G^r_{b\sa}(\va),
\eea
where $G^r_{b\sa}(\va)$ is the retarded $b$-state Green function, which, by symmetry and degeneracy, is homogeneous for all dots. \

For delocalized states, from Eq.~(\ref{eq:Hm03}) we see that $p$ is a good quantum number. Thus, the effect of coupling only leads to a shift in band energy, i.e. $\va_{dp\sa}$ becomes
$\va_{dp\sa}+{\rm Re}\Sigma_d+i \Gamma_{d\sa}(\va)/2$. We then have
\bea
I_d&=&\frac{-e}{\hbar}\sum_\sa\sum_{p=1}^{N}\int_{-W}^W\frac{d\va}{\pi}\frac{\Gamma_{L d\sa}(\va)\Gamma_{Rd\sa
}(\va)}{\Gamma_{L d\sa}(\va)+\Gamma_{Rd\sa
}(\va)}\nn\\&& \times[f_L(\va)-f_R(\va)]{\rm Im}G^r_{p\sa}(\va),
\eea
where $G^r_{p\sa}(\va)$ is the retarded $d$-state Green function. \

Because the band width of the metallic leads is commonly much larger than the other energy scales in quantum dot systems, the tunnelling rates $\Gamma_{\alpha l \sa}(\va)$ vary smoothly with energy. By assuming these rates to be energy- and spin-independent, we denote in a shorthand $\Gamma_{\alpha l \sa}(\va)=\Gamma_{\alpha l}~(l=d,b)$. To further facilitate analysis, we consider the system with a right-lead bias setting $\mu_R$ lower than the local density of states of the QDA. In such case, we can safely set $f_R(\va)=0$. We then find
\bea
I_b &\simeq&\frac{-Ne}{\hbar}\sum_\sa \int_{-W}^W\frac{d\va}{\pi}\frac{\Gamma_{Rb
}\Gamma_{L b }f_L(\va)}{\Gamma_b}{\rm Im}G^r_{b\sa}(\va)\nn\\
&=&\frac{Ne\Gamma_{Rb
}}{\hbar}\sum_\sa\lan n_{b\sa}\ran\label{eq:IandNb}.
\eea
 The tunnelling current $I_b$ and the $b$-orbital occupation number ($ \sum_\sa\lan n_{b\sa}\ran$) differ by a constant multiplicative factor $eN\Gamma_{Rb}/\hbar$. To study bistability in the tunnelling current, it suffices to compute the $b$-orbital occupation number as a function of bias voltage. The tunnelling current $I_d$ and the average $d$-orbital occupation number $\lan n_{d\sa}\ran (\equiv N^{-1}\sum_{p=1}^N\lan n_{p\sa}\ran)$ can be related in a similar fashion,
\be I_d \simeq\frac{Ne\Gamma_{Rd}}{\hbar}\sum_\sa\lan n_{d\sa}\ran. \label{eq:IandNd} \ee

\section{Green Functions\label{sec:GFs}}

To obtain the average $b$-orbital and $d$-orbital occupation numbers for tunnelling currents $I_b$ (\ref{eq:IandNb}) and $I_d$ (\ref{eq:IandNd}), we used the nonequilibrium Green function method developed by Kuo and Chang\cite{KuoChang07,Chang08}. This method keeps the self-energy from the lead-dot tunnelling coupling to lowest order, precisely corresponding to the scheme 2 of Ref.~\onlinecite{Pals1996}. Similar to the Hartree-Fock approximation\cite{Lacroix81,Hewson66}, we keep the self-energy term that describes the scattering of a conduction electron by the electron in the dots, but neglect the spin-flip scattering processes responsible for the Kondo correlations\cite{Lacroix81,Hewson66}. The important difference is that our method preserves the sum rule of the spectral density: $-\int d\va{\rm Im}G^r/\pi=1$. {Such approximation scheme is generally considered valid in the Coulomb blockade regime, in which the lead-dot coupling is too weak for the Kondo effect to be observed\cite{Pals1996,Palacios1997}. Note that our lead-dot coupling in this Green function method is not limited to the weak-coupling limit, i.e.,. $\Gamma\ll k_BT$, which the standard Master equation methods generally assume\cite{Altshuler1991,Beenakker1991}. The derivation of the Master equation in terms of the Keldysh Green
function method has been reported\cite{Schoeller1994}.} For the reasons mentioned in the previous section, we also neglect contributions from off-diagonal elements. In a wide-band limit, the self-energy term gives
\bea
\Sigma_b(\va)&=&\sum_{\alpha\bk}\frac{|t_{\alpha b\sa}|^2}{\va-\va_{\bk}}\simeq i\Gamma_b/2,\\
\Sigma_d(\va)&=&\sum_{\alpha p\eta}\frac{|t_{\alpha d\sa}|^2}{\va-\va_{p\eta}}\simeq i\Gamma_d/2,
\eea
where $\Gamma_l=\Gamma_{Ll}+\Gamma_{Rl}$ ($l=d,b$).
The tunnelling rate $\Gamma_l$, being small, produces effectively a broadening width to the discrete energy levels for the quantum dots. \

It is interesting to note that the same Green function solution can be obtained by first solving the isolated QDA system and then replacing the positive infinitesimal imaginary part $\delta$ in frequency by a finite decay width $\Gamma/2$ in a phenomenological way\cite{Haugtextbook}. \

We first derive the retarded Green functions. In the Zubarev notation\cite{Zubarev60}, a retarded Green function involving arbitrary fermionic operators $A$ and $B$ reads
\bea
\bra A , B \ket = - i \; \lim_{\delta \rightarrow 0^+} \; \int^\infty_{0} dt \; e^{i(\va + i\delta)t} \; \langle\{A(t),B(0)\}\rangle  , \label{Zubarov1}
\eea
where $\lan \cdots\ran$ denotes grand ensemble average and $\delta$ a positive infinitesimal. Using the Heisenberg equation of motion, one can show
\bea
(\va+i\delta)\langle\langle A,B\rangle\rangle = \langle\{A,B\}\rangle +\langle\langle [ A,H ] ,B\rangle\rangle \label{eq:Zubarov} .
\eea
This relation allows one to derive a flow of equations for the retarded Green functions.

Since the $|nb\sa\ran$ states are homogeneous and degenerate for all dots due to symmetry, we denote by $G^r_{b\sa}(\va)=\langle\langle b_{\sa},b^\dag_{\sa}\rangle\rangle$ and $G^r_{bb\sa}(\va)=\langle\langle n_{b\bsa} b_{\sa},b^\dag_{\sa}\rangle\rangle$ the retarded $b$-state Green functions, with the site index $n$ suppressed ($\bsa$ denotes the opposite spin of $\sa$.); for retarded $d$-state Green functions, we specifically denote $G^r_{p\sa}(\va)=\langle\langle
d_{p\sa},d^\dag_{p\sa}\rangle\rangle$, and $G^r_{pp\sa}(\va)=\langle\langle
n_{p\bsa}d_{p\sa},d^\dag_{p\sa}\rangle\rangle$, since the $|pd\sa\ran$ band states are non-degenerate. We follow the procedure of refs.~\onlinecite{KuoChang07} and \onlinecite{Chang08}. First we obtain two Green functions for the $|nb\sa\ran$ state when all the $|pd\sa\ran$ band states (total $2N$ for $N$ dots) as well as the $|(n\pm1)b\sa\ran$ states are occupied:
\be
G^r_{2N,b\sa}(\va)=\frac{(1-N_{b\bsa})c_b^2\prod_p c_p}{\mu_{b}-4\widetilde U_{xbb}-2\widetilde U_{db}}+\frac{N_{b\bsa}c_b^2\prod_p c_p}{\mu_{b}-U_{bb}-4\widetilde U_{xbb}-2\widetilde U_{db}},
\ee
\be
G^r_{2N+1,b\sa}(\va)=\frac{N_{b\bsa}c_b^2\prod_p c_p}{\mu_{b}-U_{bb}-4\widetilde U_{xbb}-2\widetilde U_{db}},
\ee
where $N_{b\bsa}=\lan n_{b\bsa}\ran$ is the single-particle occupation number in the $|b\sa\ran$ state. $c_p=\lan n_{p\sa}n_{p\bsa}\ran$ is the correlation function between $|pd\sa\ran$ and $|pd\bsa\ran$ band states, while $c_b=\lan n_{b\sa}n_{b\bsa}\ran$ is the correlation function between $|b\sa\ran$ and $|b\bsa\ran$ states; $\mu_b=\va-\va_{b\sa}+i\Gamma_b/2$.

Next, the one-particle and two-particle Green functions for b orbitals are expressed as
\bea
G^r_{b\sa}(\va)&=&(\hat{a}^\prime_b+\hat{b}^\prime_b+c_b)^2\prod_p (\hat{a}_p+\hat{b}_p+c_p)\frac{G_{2N,b\sa}}{c_b^2\prod_p c_p}\nn\\
&=&\sum_{i=1}^{5}\sum_{m=1}^{3^N}{\bar p_i}p_m\Big[\frac{(1-N_{b\bsa})}{\mu_{b}-\Pi_{xbb,i}-\Pi_{db,m}}\nn\\
&&+\frac{N_{b\bsa}}{\mu_{b}-U_{bb}-\Pi_{xbb,i}-\Pi_{db,m}}\Big]\label{eq:Gbsa},
\eea
\bea
G^r_{bb\sa}(\va)&=&(\hat{a}^\prime_b+\hat{b}^\prime_b+c_b)^2\prod_p (\hat{a}_p+\hat{b}_p+c_p)\frac{G_{2N+1,b\sa}}{c_b^2\prod_p c_p}\nn\\
&=&\sum_{i=1}^{5}\sum_{m=1}^{3^N}\frac{N_{b\bsa}{\bar p_i}p_m}{\mu_{b}-U_{bb}-\Pi_{xbb,i}-\Pi_{db,m}}\label{eq:Gbbsa}.
\eea
The above operators and parameters are defined according to Refs.~\onlinecite{KuoChang07} and \onlinecite{Chang08}: The operators $\hat{a}_p$ and $\hat{b}_p$ are carried out explicitly in Eqs.~(\ref{eq:Gbsa},\ref{eq:Gbbsa}). Specifically, they are defined such that $\hat{a}_p(f/g)=a_pf/(g+2\widetilde U_{db}/N)$, $\hat{b}_p(f/g)=b_pf/(g+\widetilde U_{db}/N)$, where
\bea
a_p&=& \lan (1-n_{p\sa})(1-n_{p\bsa})\ran=1-N_{p\sa}-N_{p\bsa}+c_p, \nn \\
b_p&=& \lan n_{p\sa}(1-n_{p\bsa})\ran+ \lan (1-n_{p\sa})n_{p\bsa}\ran\nn\\&&=N_{p\sa}+N_{p\bsa}-2c_p.\nn
\eea
Similarly, the operators $\hat{a}^\prime_b$ and $\hat{b}^\prime_b$ are defined such that $\hat{a}^\prime_b(f/g)=a_bf/(g+2\widetilde U_{xbb})$ and $\hat{b}^\prime_b(f/g)=b_bf/(g+\widetilde U_{xbb})$. The parameters $a_p,~b_p,~c_p$ ($a_b,~b_b,~c_b$) are the probability weighting factors in the Green functions for the empty, singly-, and doubly-occupied configurations in the $|pd\sa\ran$ ($|b\sa\ran$) states, respectively. $\Pi_{xbb,i}(=(i-1)\widetilde U_{xbb})$ denotes the interdot Coulomb repulsion felt by the $|b\sa\ran$ state from its neighboring $b$ orbitals in configuration $i$, while $\Pi_{db,m}$ denotes the total Coulomb repulsion felt in configuration $m$, which depends on the number of electrons occupying the $|pd\sa\ran$ states---Physically, the Green functions (\ref{eq:Gbsa},\ref{eq:Gbbsa}) compute the linear superposition of the probability amplitudes of an electron propagating in the eigenstates of the isolated QDA system. To understand detailed derivation of Green functions, readers are encouraged to see clear demonstration and discussion for the simple case of a nanostructure with two levels in Appendix. A of Ref.~\onlinecite{Chang08}.\

The probability factors $\bar p_i$'s are: $\bar p_1=a_b^2,~\bar p_2=2a_b b_b,~\bar p_3=(b_b^2+2a_b c_b),~\bar p_4=2b_b c_b,~\bar p_5=c_b^2$. In principle, Eqs.~(\ref{eq:Gbsa}) and (\ref{eq:Gbbsa}) allow one to compute the probability factors $p_m$ in terms of the $a_p,~b_p$ and $c_p$ parameters for all $3^N$ terms via a simple recursion relation. In practice, when $N$ is large, the computation becomes impractical since it scales exponentially with $N$. Fortunately, since we are considering a ring of quantum dots in which the $|b\sa\ran$ states feel the same Coulomb energy $\widetilde U_{db}/N$ to all the $|pd\sa\ran$ states, we can reduce the summation of $3^N$ terms to $2N+1$ terms, because many terms will have the same denominator, and the computation scales linearly with $N$. This observation immensely reduces the numerical complexity. The resulting Green functions read
\bea
G^r_{b\sa}(\va)=\sum_{i=1}^{5}\sum_{k=1}^{2N+1}{\bar p_i}p^\prime_k\Big[\frac{1-N_{b\bsa}}{\mu_{b}-\Pi_{xbb,i}-\Pi^\prime_{db,k}}\nn\\+\frac{N_{b\bsa}}{\mu_{b}-U_{bb}-\Pi_{xbb,i}-\Pi^\prime_{db,k}}\Big],
\eea
\bea
G^r_{bb\sa}(\va)=\sum_{i=1}^{5}\sum_{k=1}^{2N+1}\frac{N_{b\bsa}{\bar p_i}p^\prime_k}{\mu_{b}-U_{bb}-\Pi_{xbb,i}-\Pi^\prime_{db,k}},
\eea
where $\Pi^\prime_{db,k}=(k-1)\widetilde U_{db}/N$ and $p^\prime_k$ is the sum of all those $p_m$'s that share the same $\Pi^\prime_{db,k}$, which can be obtained by using the following recursion relations:
\begin{subeqnarray}
p^\prime_k (i) &=& a_i p^\prime_k (i-1) + b_i p^\prime_{k-1} (i-1)\nn\\&&+ c_i p^\prime_{k-2} (i-1); \; k<2i, \nn \\
p^\prime_k (i) &=&  b_i p^\prime_{k-1} (i-1) + c_i p^\prime_{k-2} (i-1); \; k=2i, \nn \\
p^\prime_k (i) &=&   c_i p^\prime_{k-2} (i-1); \; k=2i+1,\nn
\end{subeqnarray}
where $p^\prime_k (i)$ denote the $p^\prime_k$ coefficients at the $i^{th}$ iteration with $p^\prime_1(0)=1$ and $p^\prime_k(i)=0$ for $k<1$. The iteration process terminates at $i=N$.\

With these retarded Green functions derived, we solve for the $b$-orbital occupation numbers $N_{b\sa}$ and $c_b$ via $\displaystyle N_{b\sa}= \int \frac{d\va}{2\pi i}G^<_{b\sa}(\va)$ and $\displaystyle c_{b}= \int \frac{d\va}{2\pi i}G^<_{bb\sa}(\va)$ respectively, where the lesser Green function in the Coulomb blockade regime is related to the retarded Green function via
\bea
G^<_{b\sa}(\va)=-2i\frac{\Gamma_{L b}f_L(\va)+\Gamma_{Rb
}f_R(\va)}{\Gamma_{b }}{\rm Im}G^r_{b\sa}(\va)\label{eq:lessGF},\\
G^<_{bb\sa}(\va)=-2i\frac{\Gamma_{L b}f_L(\va)+\Gamma_{Rb
}f_R(\va)}{\Gamma_{b }}{\rm Im}G^r_{bb\sa}(\va)\label{eq:lessGF1}.
\eea
We then iterate to self-consistency. Eq.~(\ref{eq:lessGF}), while ``exact'' in the noninteracting model, {also works well for the Anderson model in the Coulomb blockade regime}. The derivation of lesser Green functions is similar to that of retarded ones. In \ref{app_lsGF}, a nanostructure system with two levels is given as an example to demonstrate Eqs.~(\ref{eq:lessGF}) and (\ref{eq:lessGF1}) by using the nonequilibrium equation of motion method\cite{Niu99}. To prove these two formulas for arbitrary levels can be done by the principle of induction, similar to the way the retarded Green function is proved\cite{Chang08}. Rigorous proof will be given elsewhere. However, from the structures of retarded and lesser Green functions in the two-level system, as well as the fact that when all Coulomb interactions are turned off, the expressions reduce to the exact solutions for noninteracting systems, it is not hard to see that Eqs.~(\ref{eq:lessGF},\ref{eq:lessGF1}) hold true for arbitrary levels.

At zero temperature, the average occupation numbers for the $|b\sa\ran$ state read
\bea
N_{b\sa}&=&\frac{\Gamma_{Lb}}{\pi\Gamma_b}\sum_{i=1}^{5}\sum_{k=1}^{2N+1}{\bar p_i}p^\prime_k\Bigg[(1-N_{b\bsa})\label{eq:occpNb}\\
&&\times\cot^{-1}\left(\frac{V-\va_{b\sa}-\Pi_{xbb,i}-\Pi^\prime_{db,k}}{\Gamma_b/2}\right)\nn\\
&&+N_{b\bsa}\cot^{-1}\left(\frac{V-\va_{b\sa}-U_{bb}-\Pi_{xbb,i}-\Pi^\prime_{db,k}}{\Gamma_b/2}\right)\Bigg],\nn
\eea
\bea
c_b&=&\frac{\Gamma_{Lb}N_{b\bsa}}{\pi\Gamma_b}\sum_{i=1}^{5}\sum_{k=1}^{2N+1}{\bar p_i}p^\prime_k\label{eq:occpcb}\\
&&\times\cot^{-1}
\left(\frac{V-\va_{b\sa}-U_{bb}-\Pi_{xbb,i}-\Pi^\prime_{db,k}}{\Gamma_b/2}\right).\nn
\eea
Besides some fixed physical parameters, it is worth noting that $N_{b\sa}$ and $c_b$ are functions of the opposite-spin occupation number $N_{b\bsa}$ and the applied bias $V$.\

Similar treatment can be used for the extended $d$-state occupation numbers $ N_{p\sa}$ and $c_p$. We have
\bea
 N_{p\sa}&=& \frac{\Gamma_{Ld}}{\pi\Gamma_d}\sum_{k=1}^{2N-1}p^\prime_k(p)\Bigg\{(1-N_{p\bsa})\\
&&\times\Bigg[a_b\cot^{-1}\left(\frac{V-\va_{dp\sa}-\Pi^\prime_{dd,k}}{\Gamma_d/2}\right)\nn\\
&&+ b_b\cot^{-1}\left(\frac{V-\va_{dp\sa}-\widetilde U_{db}-\Pi^\prime_{dd,k}}{\Gamma_d/2}\right)\nn\\
&&+ c_b\cot^{-1}\left(\frac{V-\va_{dp\sa}-2\widetilde U_{db}-\Pi^\prime_{dd,k}}{\Gamma_d/2}\right)\Bigg]\nn\\
&&+N_{p\bsa}\Bigg[ a_b\cot^{-1}\left(\frac{V-\va_{dp\sa}-\Pi^\prime_{dd,k+1}}{\Gamma_d/2}\right)\nn\\
&&+ b_b\cot^{-1}\left(\frac{V-\va_{dp\sa}-\widetilde U_{db}-\Pi^\prime_{dd,k+1}}{\Gamma_d/2}\right)\nn\eea
\bea
&&+ c_b\cot^{-1}\left(\frac{V-\va_{dp\sa}-2\widetilde U_{db}-\Pi^\prime_{dd,k+1}}{\Gamma_d/2}\right)\Bigg]\Bigg\},\nn
\eea
\bea
c_p&=&\frac{\Gamma_{Ld}N_{p\bsa}}{\pi\Gamma_d}\sum_{k=1}^{2N-1}p^\prime_k(p)\Bigg[\\
&&a_b\cot^{-1}\left(\frac{V-\va_{dp\sa}-\Pi^\prime_{dd,k+1}}{\Gamma_d/2}\right)\nn\\
&&+ b_b\cot^{-1}\left(\frac{V-\va_{dp\sa}-\widetilde U_{db}-\Pi^\prime_{dd,k+1}}{\Gamma_d/2}\right)\nn\\
&&+ c_b\cot^{-1}\left(\frac{V-\va_{dp\sa}-2\widetilde U_{db}-\Pi^\prime_{dd,k+1}}{\Gamma_d/2}\right)\Bigg],\nn
\eea
where $\Pi^\prime_{dd,k}=(k-1)\widetilde U_{dd}/N$. $p^\prime_k(p)$ is calculated in a similar way to $p^\prime_k$ except that the operator $(\hat{a}_p+\hat{b}_p+c_p)$ is removed. The effect by the operator $(\hat{a}_b+\hat{b}_b+c_b)$ is carried out explicitly when calculating $G_{p\sa}(\va)$ and $G_{pp\sa}(\va)$. Note that $\hat{a}_b$ and $\hat{b}_b$ operators have different definitions from the $\hat{a}^\prime_b$ and $\hat{b}^\prime_b$ operators: $\hat{a}_b(f/g)=a_bf/(g+2\widetilde U_{db})$ and $\hat{b}_b(f/g)=b_bf/(g+\widetilde U_{db})$.\

To obtain the occupation numbers at nonzero temperatures $T$, we first define a function $F(x,y)$
\bea
F(x,y)&=&\frac{\pi}{2}-{\rm Im} \left[\Psi\left(\frac{1}{2}+\frac{V-x+iy}{2\pi i k_BT}\right)\right],
\eea
where $\Psi$ denotes the digamma function. Thus, at nonzero temperatures, the average occupation numbers for the $|b\sa\ran$ state read
\bea N_{b\sa}&=&\frac{\Gamma_{Lb}}{\pi\Gamma_b}\sum_{i=1}^{5}\sum_{k=1}^{2N+1}{\bar p_i}p^\prime_k\Bigg[(1-N_{b\bsa})\\
&&\times F(\va_{b\sa}+\Pi_{xbb,i}+\Pi^\prime_{db,k},\Gamma_b/2)\nn\\
&&+N_{b\bsa}F(\va_{b\sa}+U_{bb}+\Pi_{xbb,i}+\Pi^\prime_{db,k},\Gamma_b/2)\Bigg]\nn,
\eea
\bea
c_b&=&\frac{\Gamma_{Lb}N_{b\bsa}}{\pi\Gamma_b}\sum_{i=1}^{5}\sum_{k=1}^{2N+1}{\bar p_i}p^\prime_k\nn\\ &&\times F(\va_{b\sa}+U_{bb}+\Pi_{xbb,i}+\Pi^\prime_{db,k},\Gamma_b/2).
\eea
Similarly for $N_{p\sa}$ and $c_p$.

\section{Results\label{sec:Results}}
It is shown from Eqs.~(\ref{eq:IandNb}) and (\ref{eq:IandNd}) that the tunnelling current as a function of bias $V$ can be related to the average occupation numbers up to a constant multiplicative factor. It therefore suffices to investigate the average $b$- and $d$-orbital occupation numbers in order to understand the transport properties of the ring of coupled quantum dots. We shall consider two kinds of limiting cases: (i) $t_{\alpha d\sa}\ll t_{\alpha b\sa}$ and (ii) $t_{\alpha b\sa}\ll t_{\alpha d\sa} $. \

The first kind can be realized in QDAs in which the smaller dot (which hosts the localized $b$ orbital) is elongated, thereby allowing the orbital wave function to stretch out in the direction perpendicular to the plane of the ring (see Fig.~\ref{fig:Conf_QDA}). It then hybridizes with the electronic states in the leads a lot stronger than the delocalized orbital. This system will be used as an example in Sec.\ref{sec:asym_rate}-Sec.\ref{sec:FiniteT} to illustrate the effects on bistability produced by all the relevant physical parameters. \

\begin{figure}[t]
 \centering
 \subfigure[] { \label{fig:DiffALR_ZN150}
\includegraphics[scale=0.62]{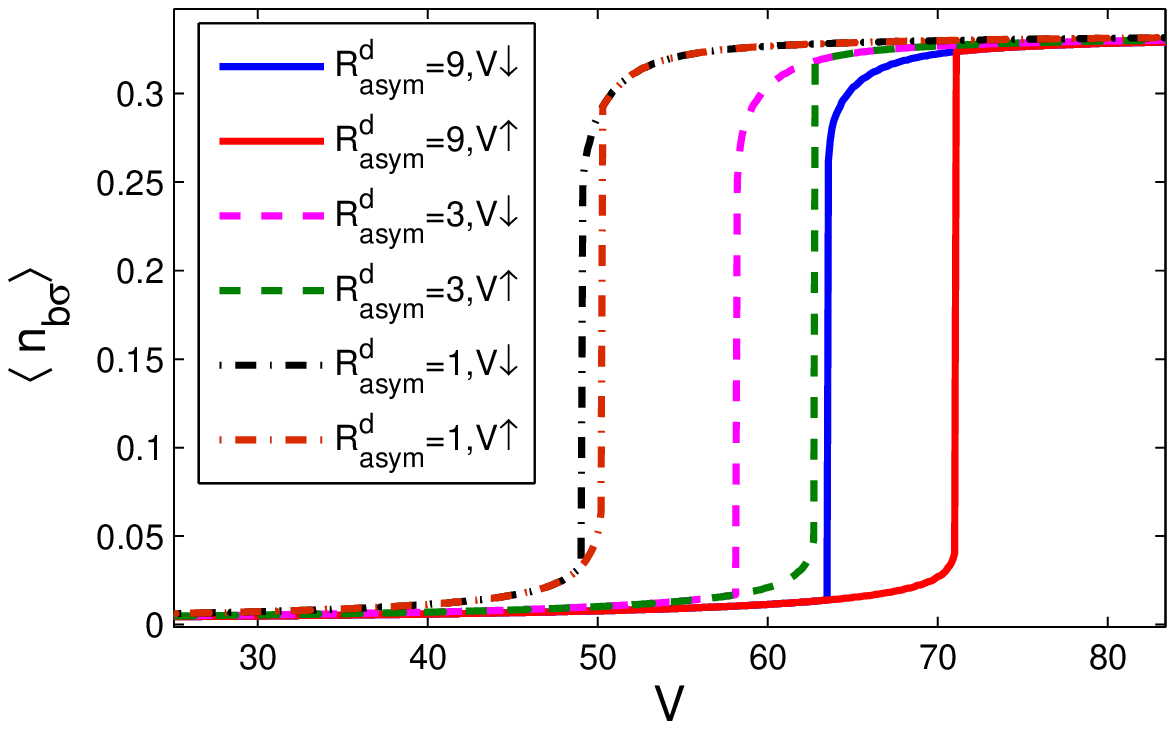}}\
\subfigure[] { \label{fig:DiffALR_XN150}
\includegraphics[scale=0.62]{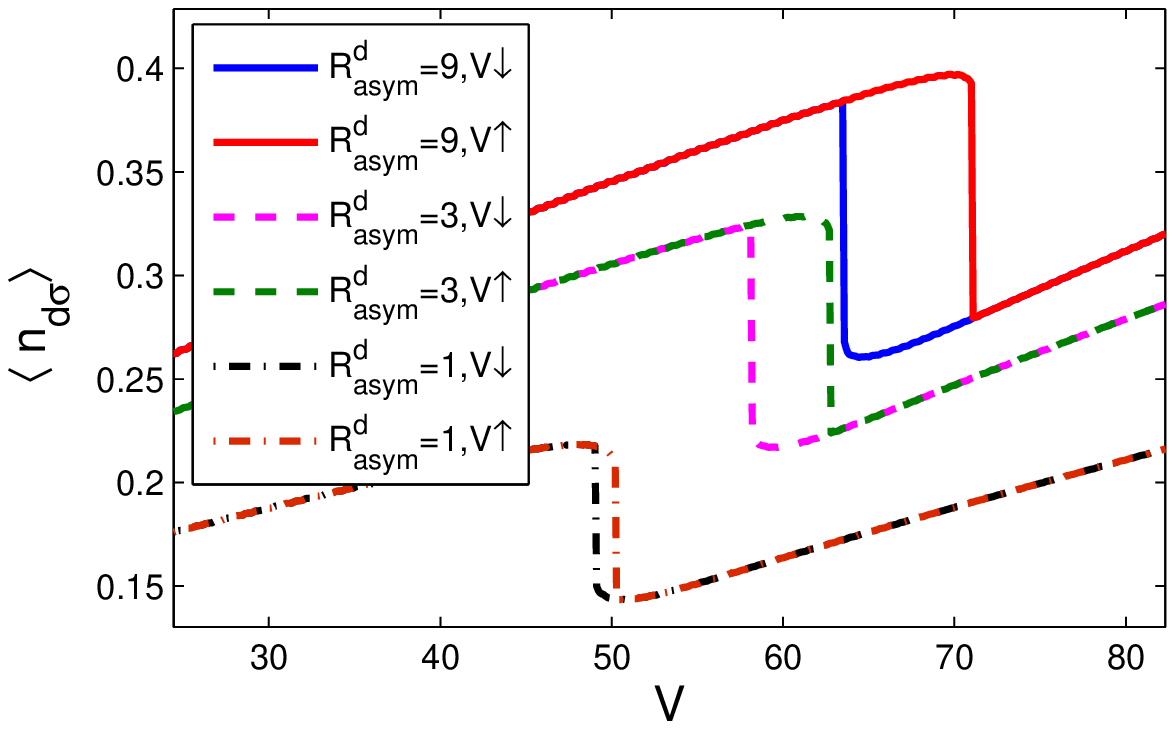} }\\
\caption{\protect\small{(Color online)(a) The average $b$-orbital occupation number $\lan n_{b\sa}\ran$ (b) The average $d$-orbital occupation number $\lan n_{d\sa}\ran ~(\equiv \sum_{p=1}^{N}\lan n_{p\sa}\ran/N)$ as a function of bias $V$ (in units of $\Gamma$) as we sweep up ($V\uparrow$) and down ($V\downarrow$) the voltage bias. Different left-right tunnelling asymmetric factors $R^d_{\rm asym}=1,~3, ~9$ are taken, where $R^d_{\rm asym}=\Gamma_{Ld}/\Gamma_{Rd}$. The curves at zero temperature $T=0$ show that the bistable effect enhances as the asymmetric factor $R^d_{\rm asym}$ increases.. We used $N=100$ dots, $U=70\Gamma,~t_d=30\Gamma,~\va_{b\sa}=\va_{d\sa}=30\Gamma,~\Gamma_b=2\Gamma$ and $\Gamma_d=0.1\Gamma$. The Coulomb energies are $U_{bb}=U$, $\widetilde U_{db}=\widetilde U_{dd}\simeq 0.84U$, and $\widetilde U_{xbb}=0$. All energy quantities are in units of a characteristic energy, $\Gamma$.
  }}\label{fig:diffALR}
\end{figure}

The second kind can be realized in QDAs in which the small dots are embedded in a core-shell structure. It is natural for this case to assume that the tunnelling rate for the delocalized orbital is much larger. The tunnelling characteristics of the second kind and its difference from the first kind will be discussed in Sec.\ref{sec:tblltd}. \

Throughout the paper, all energy quantities are in units of a characteristic energy, $\Gamma$, which depends on systems of interest. Furthermore, we take $\Gamma_{Lb}=\Gamma_{Rb}=\Gamma_b/2$, while considering various ratios of left and right tunnelling rates for the $d$-state, $R^d_{\rm asym} = \Gamma_{Ld}/\Gamma_{Rd}$, as discussed in Sec.\ref{sec:asym_rate}. Varying the ratio $\Gamma_{Lb}/\Gamma_{Rb}$ produces similar effect.\

For the Coulomb interactions (\ref{eq:tran_H05}), we first set $U_{bb}=U$, which will also be used as a characteristic Coulomb energy. We then take interdot and intradot Coulomb interactions to be $U_{xjl}= U_{jl}/2$ and $U_{db}=U_{dd}=U/4$, respectively. This will give $\widetilde U_{db}=\widetilde U_{dd}\simeq U[1+(\gamma+{\rm ln}2N/\pi)/2]/4$ without the screening effect. To clearly demonstrate the influences of the system parameters, particularly $U$ and the QDA size $N$, over bistability in the QDA system,  we will first focus on the ideal system, i.e., QDA without the inter $b$-orbital Coulomb interaction and screening effect by setting $\widetilde U_{xbb}=0$ and $L_{jl}\rightarrow\infty$. These two effects will be considered in Sec.\ref{sec:Uxdd}.

To understand bistability in the coupled dots systems, it is instructive to recognize its link with the population switching between the localized and delocalized states: Collectively, due to a large tight-binding parameter $t_d$, the $d$ states form a finite band. The states in the lower part of the band get first populated as we increase the bias, see Fig.~\ref{fig:Conf_QDA}(d). Consequently, the energy level of the $b$ states gets pushed up by the Coulomb repulsion to higher energy, therefore postponing the occurrence of its population, till those $d$ states whose energies are below the {\it effective} energy level of the $b$ states are all filled. Right as we continue increasing the bias when these $d$ states are filled, a population switching between some higher-energy $d$ states and the $b$ states occurs. It results in a sudden jump in the tunnelling current as well as the average $b$-orbital occupation number, and a sudden fall in the average $d$-orbital one.\

Reversely, as we decrease the bias from high-bias, we start to deplete the $d$ states, which results in pushing the $b$-state energy level downward, till those $d$ states whose energies are above the {\it effective} energy level of the $b$ states are all depleted, then a population switching occurs. Bistability shows up when these two events of switching occur at different biases when we sweep upward and downward the bias voltage. Note, however, that the signature of population switching does not guarantee bistability, as will be shown in Sec.\ref{sec:CoulombU}. We emphasize that the mechanism of population switching discussed here is to be distinguished in physical nature from that of the two-level quantum dots\cite{golosov2007} or the likes.\

\begin{figure}[t]
 \centering
 \subfigure[] { \label{fig:N150Z_comparison}
\includegraphics[scale=0.64]{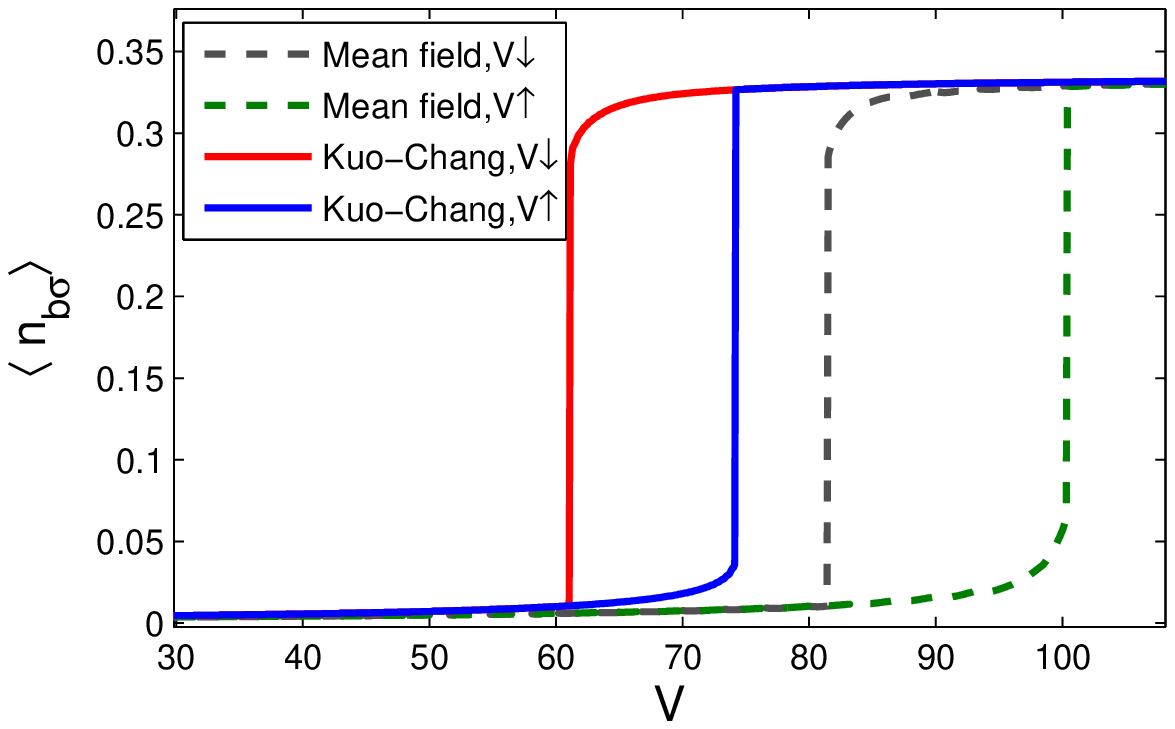} }\
 \subfigure[] { \label{fig:N150X_comparison}
\includegraphics[scale=0.64]{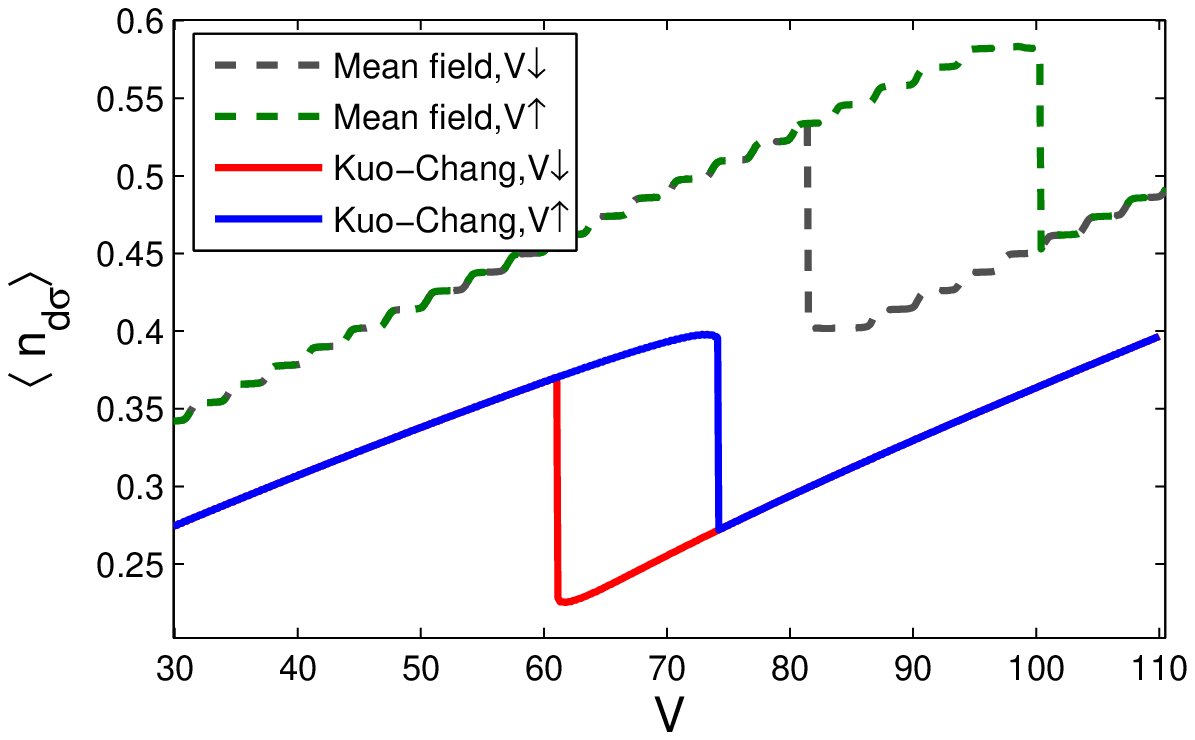}}\\
\caption{\protect\small {(Color online)(a) The $b$-orbital occupation number $\lan n_{b\sa}\ran$ (b) The average $d$-orbital occupation number $\lan n_{d\sa}\ran$ as a function of bias $V$ (in units of $\Gamma$) obtained by the mean-field theory and Kuo-Chang method\cite{KuoChang07,Chang08} for a ring of coupled dots with  $N=150$, with $R^d_{\rm asym}=9$. The other parameters are the same as in Fig.~\ref{fig:diffALR}.  }}\label{fig:N150_comparison}
\end{figure}

While discussing in Sec.\ref{sec:asym_rate} and Sec.\ref{sec:CoulombU} some of the results that were already shown by a simple mean-field approach\cite{KuoChang09}, we stress that our more complex method focuses mainly on the temperature and size effects of the QDA system. Comparison between these two methods is made in Sec.\ref{sec:Comp_MFT}.

\subsection{Effect of asymmetry of left and right tunnelling rates\label{sec:asym_rate}}

From the above viewpoint of population switching, it is not surprising that the bistable effect is enhanced as we increase the $d$-orbital occupation numbers by increasing the degrees of freedom that form a band, like extending from 1D to 2D QDA or by tuning the left and right tunnelling rates. The latter effect on bistability is demonstrated in Fig.~\ref{fig:diffALR} in the limit $t_{\alpha d\sa}\ll t_{\alpha b\sa}$, where the asymmetric factor $R^d_{\rm asym}=\Gamma_{Ld}/\Gamma_{Rd}$ is varied. The figure shows that as $R^d_{\rm asym}$ increases, so does the average $d$-orbital occupation number. As a result, the bistable effect enhances. \

\begin{figure}[t]
 \centering
 \subfigure[] { \label{fig:ALR9to1_ZdiffN}
\includegraphics[scale=0.65]{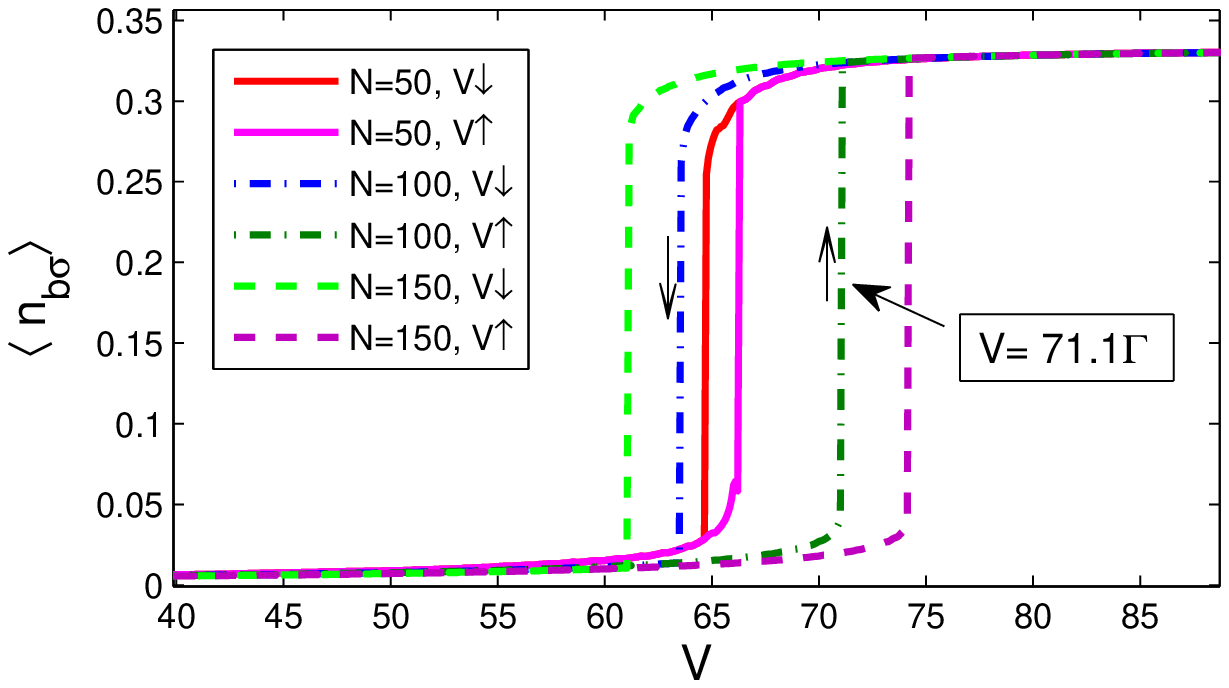} }\
\subfigure[] { \label{fig:ALR9to1_XdiffN}
\includegraphics[scale=0.65]{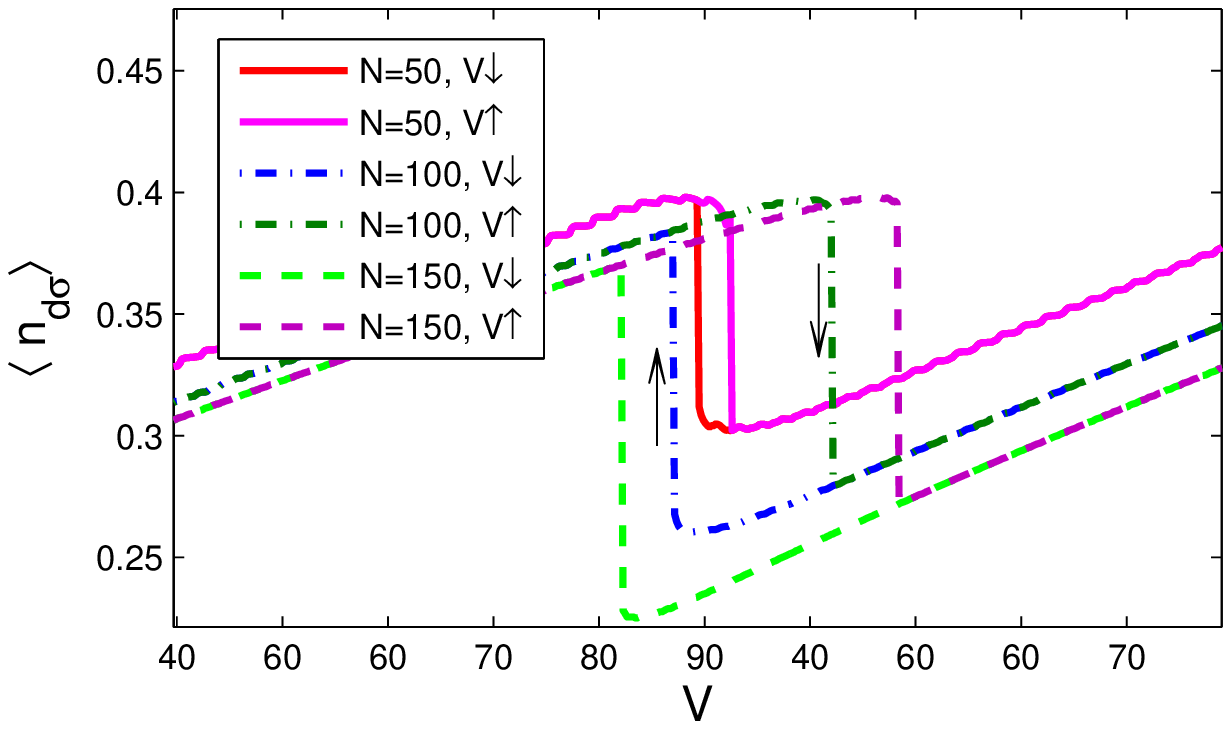}}\\
\caption{\protect\small{(Color online)(a) The average $b$-orbital occupation number $\lan n_{b\sa}\ran$ (b) The average $d$-orbital occupation number $\lan n_{d\sa}\ran$ as a function of bias $V$ (in units of $\Gamma$) for a ring of $N$ coupled dots with $N=50,~100,~150$, with $R^d_{\rm asym}=9$. $\widetilde U_{db}=\widetilde U_{dd}\simeq 0.75U,~0.84U$, and $0.89U$ for $N=50,~100$, and $150$ respectively. As the size number $N$ of dots increases, bistability enhances accordingly. Rest of parameters are same as in Fig.~\ref{fig:diffALR}.  }}\label{fig:diffN}
\end{figure}

\subsection{Comparison with mean-field theory\label{sec:Comp_MFT}}
It is of interest to compare the solutions obtained from two different methods: the mean-field approach\cite{KuoChang09} and the Kuo-Chang method\cite{KuoChang07,Chang08}. It is well known that mean-field theory breaks automatically the particle-hole symmetry of the system Hamiltonian. Moreover, being an effective single-particle approximation, it does not treat two-particle correlation adequately, particularly when the correlation is strong. Recently, the nonequilibrium mean-field theory has been criticized for possibly not being valid and giving false results, regarding bistability or hysteresis\cite{Horvath2008}. The present work, using a more careful treatment of finite-size quantum dot systems, validates in some respects the mean-field results that could otherwise raise doubt.\

Instead of treating two-particle correlation in a self-consistent way like the Kuo-Chang method\cite{KuoChang07,Chang08}, in mean-field theory, we decouple two-particle Green functions between the localized $b$ and delocalized $d$ states, and those among the $d$ states in the equations of motion into a product of a single-particle occupation number and a single-particle Green function. Namely, we make the following approximations
\begin{subeqnarray}
\langle\langle n_{p\sp}b_{\sa},b^\dag_{\sa}\rangle\rangle&\approx&\lan n_{p\sp}\ran \langle\langle b_{\sa},b^\dag_{\sa}\rangle\rangle,\\
\langle\langle
n_{b\sp}d_{p\sa},d^\dag_{p^\prime\sa}\rangle\rangle&\approx&\langle
n_{b\sp}\rangle \langle\langle d_{p\sa},d^\dag_{p^\prime\sa}\rangle\rangle,\\
\langle\langle
n_{p\bsa}d_{p\sa},d^\dag_{p^\prime\sa}\rangle\rangle&\approx&\langle
n_{p\bsa}\rangle \langle\langle d_{p\sa},d^\dag_{p^\prime\sa}\rangle\rangle.
\end{subeqnarray}
Since we expect the Coulomb interactions between these states are suppressed by a factor of $N$ (the total number of dots), we expect the mean-field results to be sensible in the large $N$ limit. Note, however, we leave $\langle\langle n_{b\bsa}b_{\sa},b^\dag_{\sa}\rangle\rangle$ intact for the localized $b$ orbitals because its correlation is expected to be strong. Details of calculation can be found in Ref.~\onlinecite{KuoChang09}.

Figure \ref{fig:N150_comparison} compares the bistability results from these two methods. The results show that the mean-field approach overestimates the bistable effect by $44\%$. The discrepancy for this effective single-particle theory is possibly attributed to artificial spontaneous symmetry breaking and incorrect decoupling of two-particle correlations under the assumption that their correlations are small.

\begin{figure}[t]
 \centering
 \subfigure[] { \label{fig:N100ALR9to1V81.6DOS}
\includegraphics[scale=0.55]{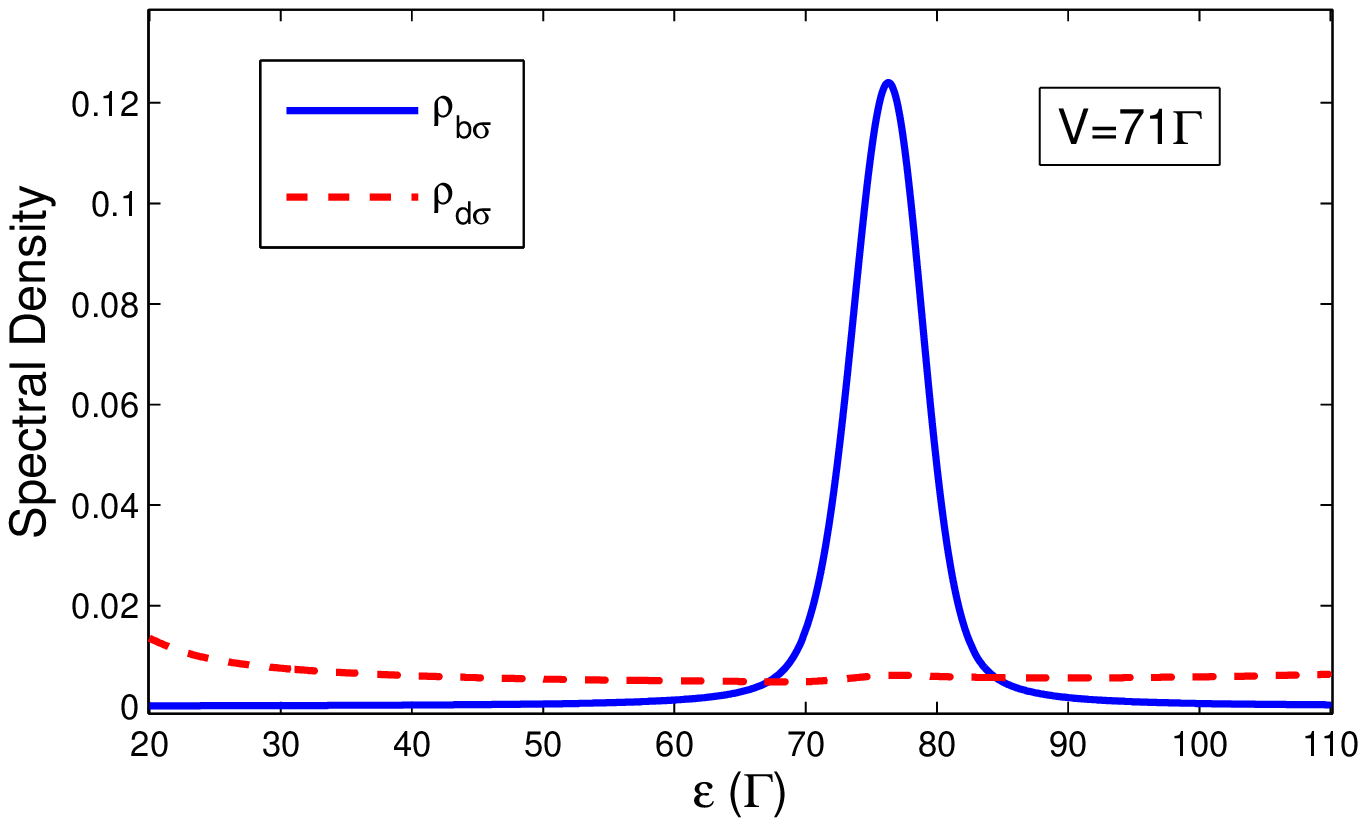} }\
  \subfigure[] { \label{fig:N100ALR9to1V82.2DOS}
\includegraphics[scale=0.55]{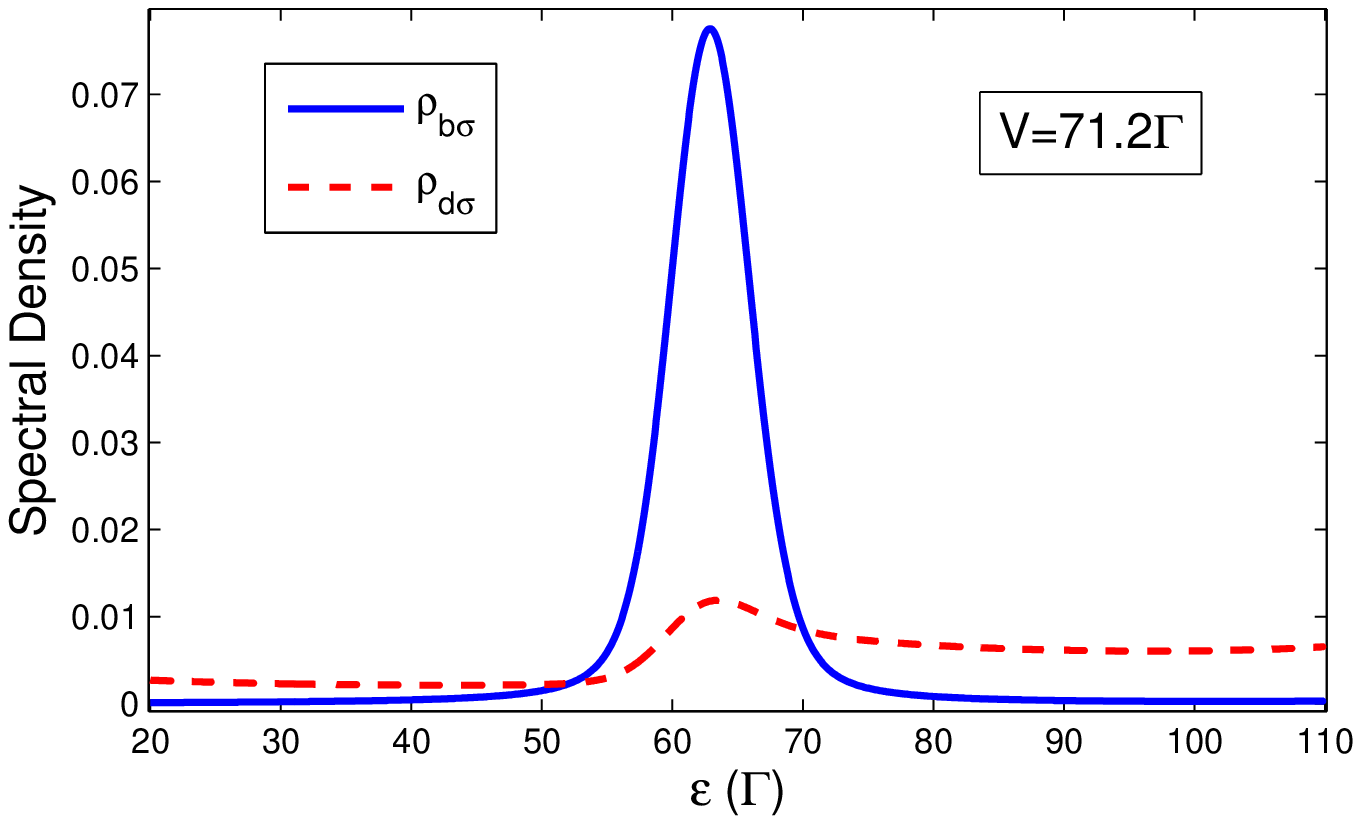}}\\
\caption{\protect\small{(Color online)The average $b$-orbital spectral density $\rho_{b\sa}(\va)=-{\rm Im}G_{b\sa}(\va)/\pi$ and the average $d$-orbital spectral density $\rho_{d\sa}(\va)=-{\rm Im}\sum_p G_{p\sa}(\va)/(\pi N)$ as a function of frequency $\va$ (in units of $\Gamma$) at bias (a) $V=71\Gamma$ (b) $V=71.2\Gamma$ in the case of increasing bias. (a) and (b) correspond to the spectral density before and after the switching point $V\simeq 71.1\Gamma$, respectively; see the dash-dotted curve in Fig.~\ref{fig:ALR9to1_ZdiffN}. In plotting $\rho_{d\sa}(\va)$, we artificially broadened the spectrum by multiplying $\Gamma_d$ by a factor of $20$. We used $R^d_{\rm asym}=9$, $N=100$. The other parameters are the same as in Fig.~\ref{fig:diffALR}.  }}\label{fig:DOS}
\end{figure}

\subsection{Effect of QDA size\label{subsec:size}}
A simple mean-field theory\cite{KuoChang09} has demonstrated that devices made of 1D and 2D QDAs can exhibit bistability in the tunnelling current as we tune the bias voltage. However, an important question remains as to how large a QDA must be in order to realize such bistability. This determines the density of memory bits made of such devices. The emergence of bistability in these systems is tightly related to the formation of a ``good'' band in one or two of the orbital states. But how do we judge if it forms a good band? In Fig.~\ref{fig:diffN}, we show for large enough tight-binding parameter $t_d$ and Coulomb energy $U$, the bistable regime (the bias interval where there exist two solutions for nonequilibrium states) starts to show up at $N\sim 50$. As we enlarge the size $N$ of the ring of coupled dots (a 1D QDA), the bistable effect enhances accordingly. This enhancement is caused both by a better band formation and by increasing $\widetilde U_{db}$ and $\widetilde U_{dd}$. \

\begin{figure}[t]
 \centering
 \subfigure[] { \label{fig:diff_tcALR9to1N100Z}
\includegraphics[scale=0.59]{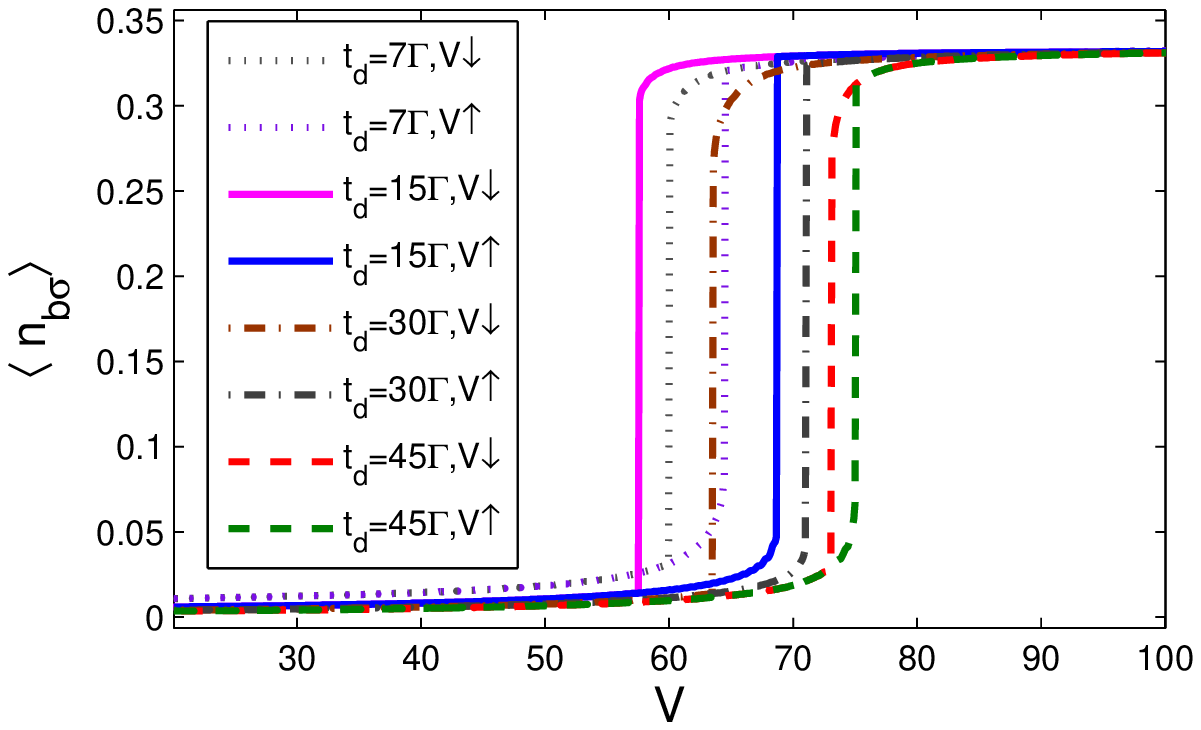} }\
 \subfigure[] { \label{fig:diff_tcALR9to1N100X}
\includegraphics[scale=0.6]{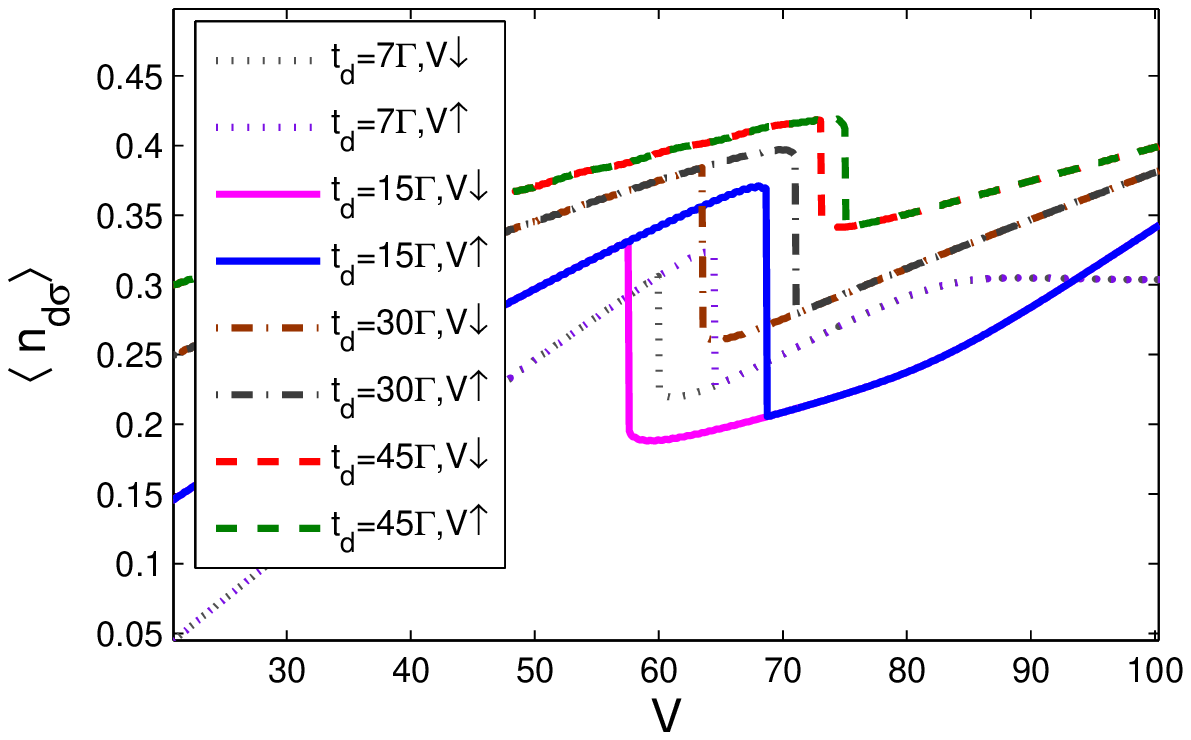}}\\
\caption{\protect\small{(Color online)(a) The average $b$-orbital occupation number $\lan n_{b\sa}\ran$ (b) The average $d$-orbital occupation number $\lan n_{d\sa}\ran$ as a function of bias $V$ (in units of $\Gamma$) for different tight-binding parameters $t_d=7\Gamma,~15\Gamma,~30\Gamma,~45 \Gamma$  for a QDA of size $N=100$, with $R^d_{\rm asym}=9$. Rest of parameters are the same as in Fig.~\ref{fig:diffALR}.  }}\label{fig:difft_d}
\end{figure}

The reason that a system of size $N\sim 100$ is enough to form a good (quasi) band can be understood by looking at the maximum energy level spacing for the $d$ states, ${\rm max}(\{\Delta_{p}\})\approx 4\pi t_d/N$ and the effective Coulomb energy $\widetilde U_{dd}/N$ felt among them. Figure \ref{fig:DOS} shows the spectral density for a ring of coupled dots with $N=100$ as a function of energy at a bias before (see Fig.~\ref{fig:N100ALR9to1V81.6DOS}) and after (see Fig.~\ref{fig:N100ALR9to1V82.2DOS}) the population switching point ($V=71.1\Gamma$), which can be seen from the dash-dotted curve in Fig.~\ref{fig:ALR9to1_ZdiffN}. It is shown that the spectral density in the $d$ states undergoes a tremendous redistribution before and after the switching point, while the peak in the spectral density of the $b$ states shifts its position to the left. These sudden redistributions in the density of states reflect the sudden increase and fall of the average $b$- and $d$-orbital occupation numbers, respectively. \

From the parameters used in Fig.~\ref{fig:DOS}, $t_d=30\Gamma$ and $U=70\Gamma$, we get ${\rm max}(\{\Delta_{p}\})\approx 3.6\Gamma$ and $\widetilde U_{dd}/N\approx 0.6\Gamma$. On the other hand, the ``effective'' $b$-state tunnelling rate $\tilde{\Gamma}_b$ estimated from Fig.~\ref{fig:DOS} gives roughly $6.4\Gamma$ and $7.4\Gamma$ before and after the switching point, both of which are a few times larger than the bare tunnelling rate $\Gamma_b (=2\Gamma)$. With both $d$-state energy level spacings and effective Coulomb energy $\widetilde U_{dd}/N$ being small, and a large effective $b$-state tunnelling rate, it is reasonable that, as far as the $b$ state is concerned, it ``sees'' a good band, which triggers large population switching and consequently makes the bistable effect possible.

\begin{figure}[t]
 \centering
 \subfigure[] {\label{fig:ALR9to1_diffU_Z}
\includegraphics[scale=0.61]{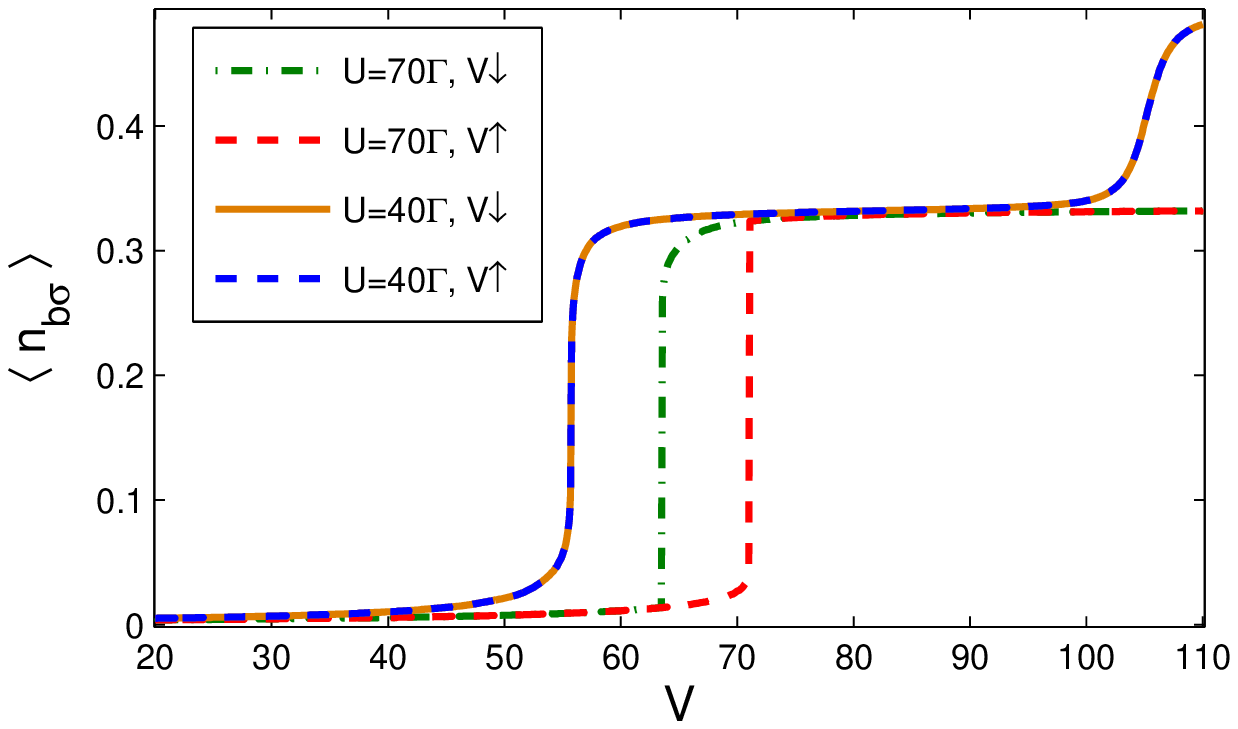} }\
  \subfigure[] { \label{fig:ALR9to1_diffU_X}
\includegraphics[scale=0.61]{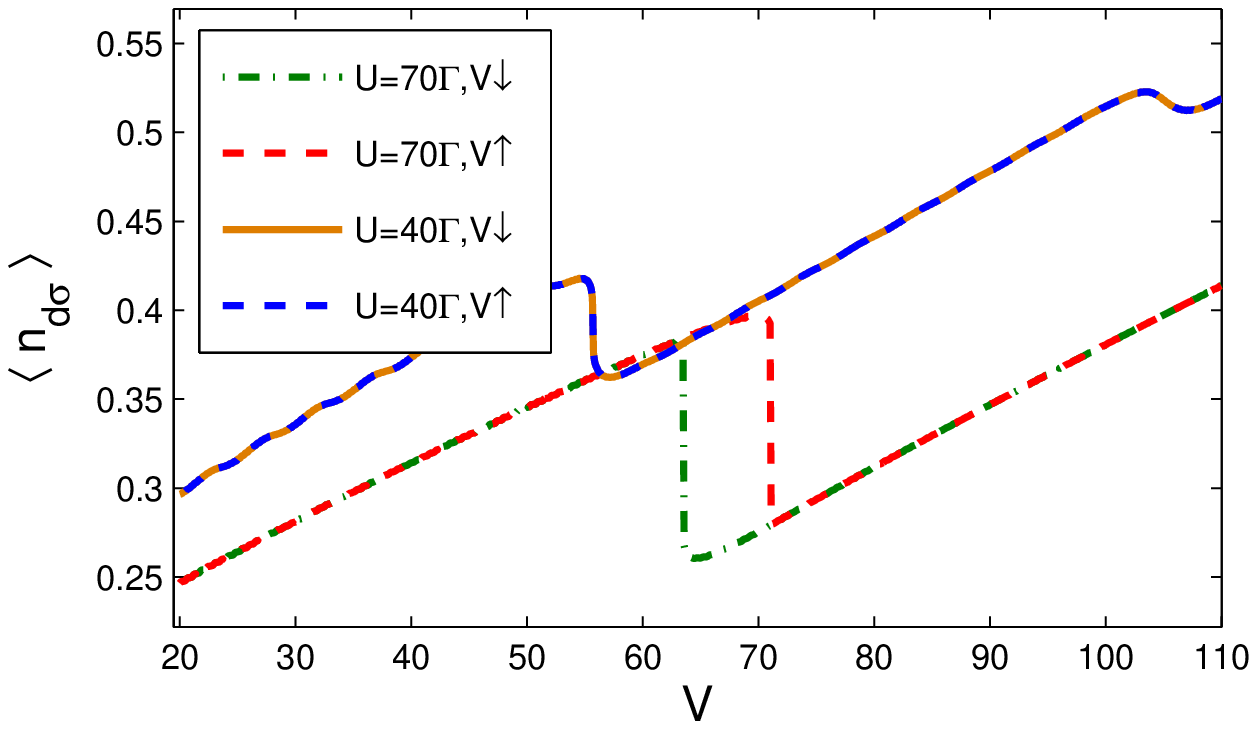}}\\
\caption{\protect\small{(Color online)(a) The average $b$-orbital occupation number $\lan n_{b\sa}\ran$ (b) The average $d$-orbital occupation number $\lan n_{d\sa}\ran$ as a function of bias $V$ (in units of $\Gamma$) for various Coulomb energy values $U=40\Gamma, ~70\Gamma$ for a ring of coupled dots with $N=100$, with $R^d_{\rm asym}=9$. The other parameters are the same as in Fig.~\ref{fig:diffALR}.  }}\label{fig:diffU}
\end{figure}

\subsection{Effect of tight-binding interaction $t_d$}
Here we investigate the dependence of bistability on the tight-binding interaction, $t_d$. The mean-field approach\cite{KuoChang09} concluded that the larger the $t_d$ value is, the more effective bistability becomes. However, this is not the case when we consider a finite-size QDA. As was discussed above, one necessary condition for bistability to take place is for the maximum  energy level spacing among the $d$ states to be sufficiently smaller than the effective $b$-state tunnelling rate. As shown in Fig.~\ref{fig:difft_d}, for a ring of coupled dots with $N=100$, increasing the $t_d$ value will certainly bring $d$-state energy level spacings to be larger than the effective $b$-state tunnelling rate (broadening energy), thus destroying the (quasi-)continuous band structure and consequently bistability. For the four cases shown in Fig.~\ref{fig:difft_d}, we estimate a suitable choice for $t_d$ value to be around $15\Gamma(\approx \widetilde U_{dd}/4)$.

\begin{figure}[t]
 \centering
 \subfigure[] {\label{fig:interdotUxbb_Z}
\includegraphics[scale=0.59]{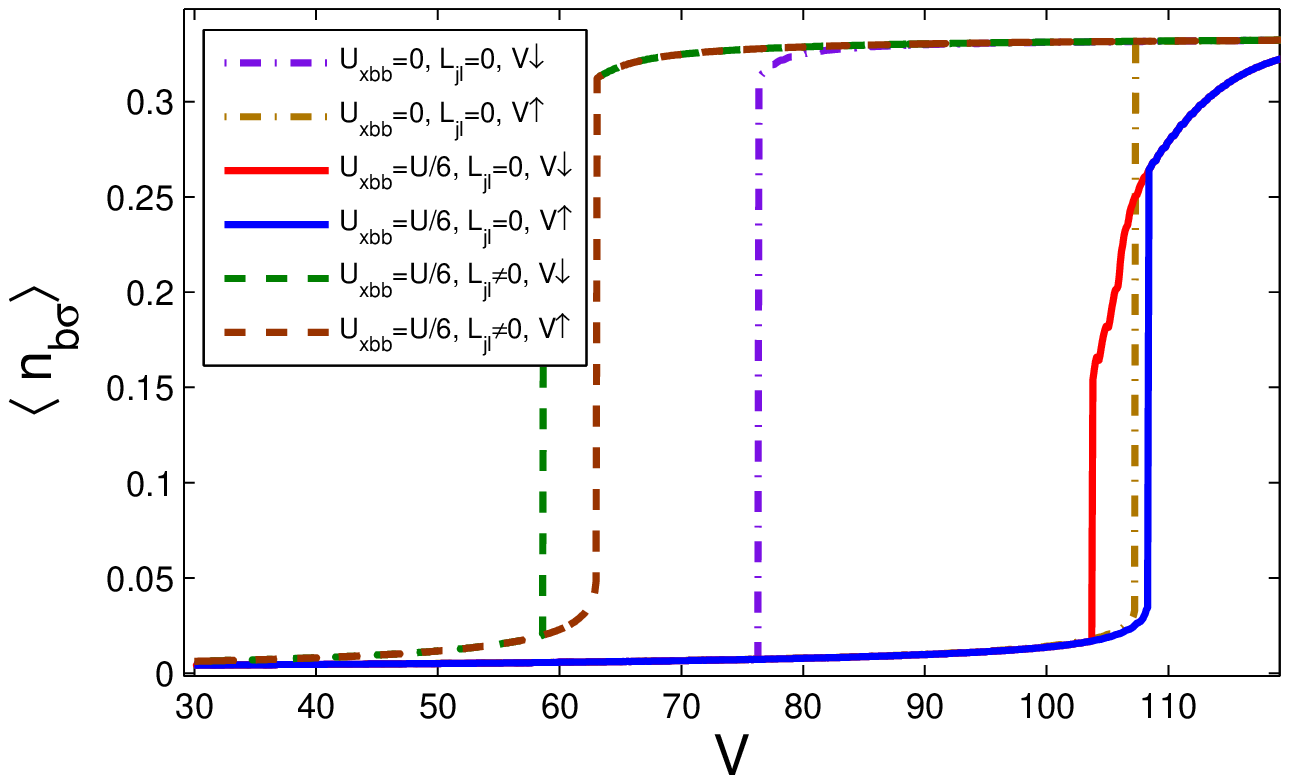} }\
  \subfigure[] { \label{fig:interdotUxbb_X}
\includegraphics[scale=0.59]{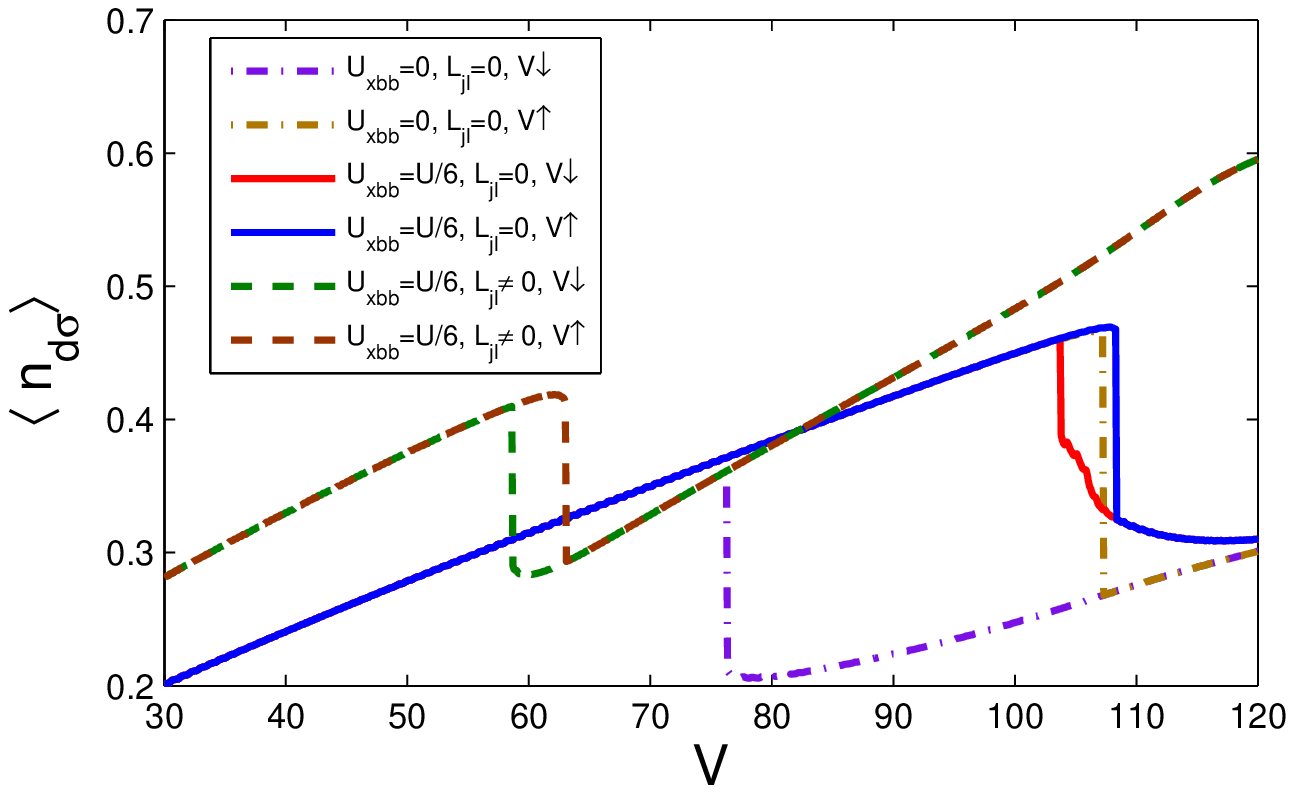}}\\
\caption{\protect\small{(Color online)(a) The average $b$-orbital occupation number $\lan n_{b\sa}\ran$ (b) The average $d$-orbital occupation number $\lan n_{d\sa}\ran$ as a function of bias $V$ (in units of $\Gamma$) for inter $b$-orbital Coulomb energy values {$U_{xbb}=0$ and $U/6$ without screening (dash-dotted, solid curves), and $ U_{xbb}=U/6$ with screening (dashed curve). The screening lengths are $L_{bb}=R_1/2,~L_{db}=2R_1/3$ and $L_{dd}=R_1$, where $R_1$ is the nearest-neighbor distance. We chose $U_{bb}=U,~U_{db}=0.6U,~U_{dd}=0.5U$ and $U_{xdb}=U_{db}/4,~U_{xdd}=U_{dd}/4$ (see the text for discussion). We also took the QDA size $N=100$, $U=70\Gamma$, $t_d=20 \Gamma$ and $R^d_{\rm asym}=9$.} The other parameters are the same as in Fig.~\ref{fig:diffALR}.  }}\label{fig:Uxbb}
\end{figure}

\subsection{Effect of Coulomb energy $U$\label{sec:CoulombU}}
A QDA of large enough size is not sufficient to realize bistability. Another condition is for the Coulomb energy $U$ to be large. As is shown in Fig.~\ref{fig:diffU}, a small $U$ does not exhibit bistability, even though there forms a good band in the $d$ states. This is simply due to the fact that for a small $U$, the $b$ state does not give the $d$ states enough leverage to push its energy level sufficiently high to trigger bistability, even though population switching starts to show up.

\begin{figure}[t]
 \centering
\subfigure[] { \label{fig:NzdiffTtc15N100}
\includegraphics[scale=0.62]{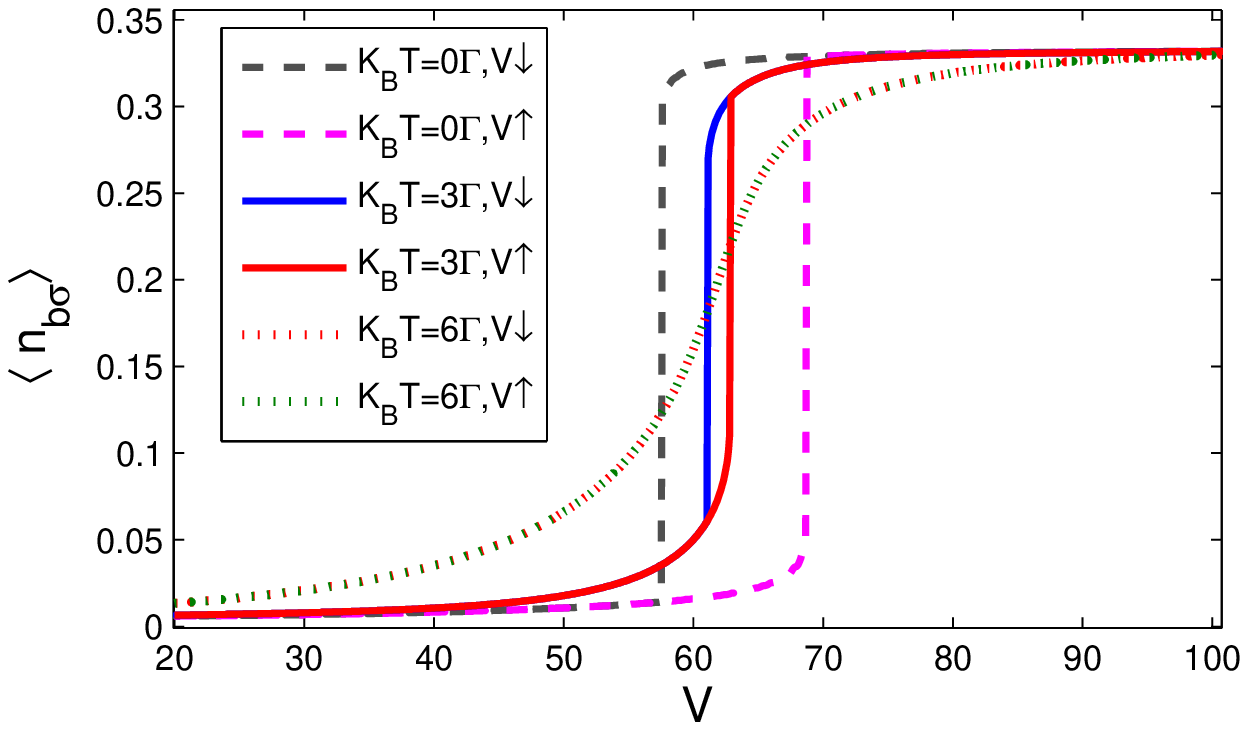} }\
 \subfigure[] { \label{fig:NxdiffTtc15N100}
\includegraphics[scale=0.62]{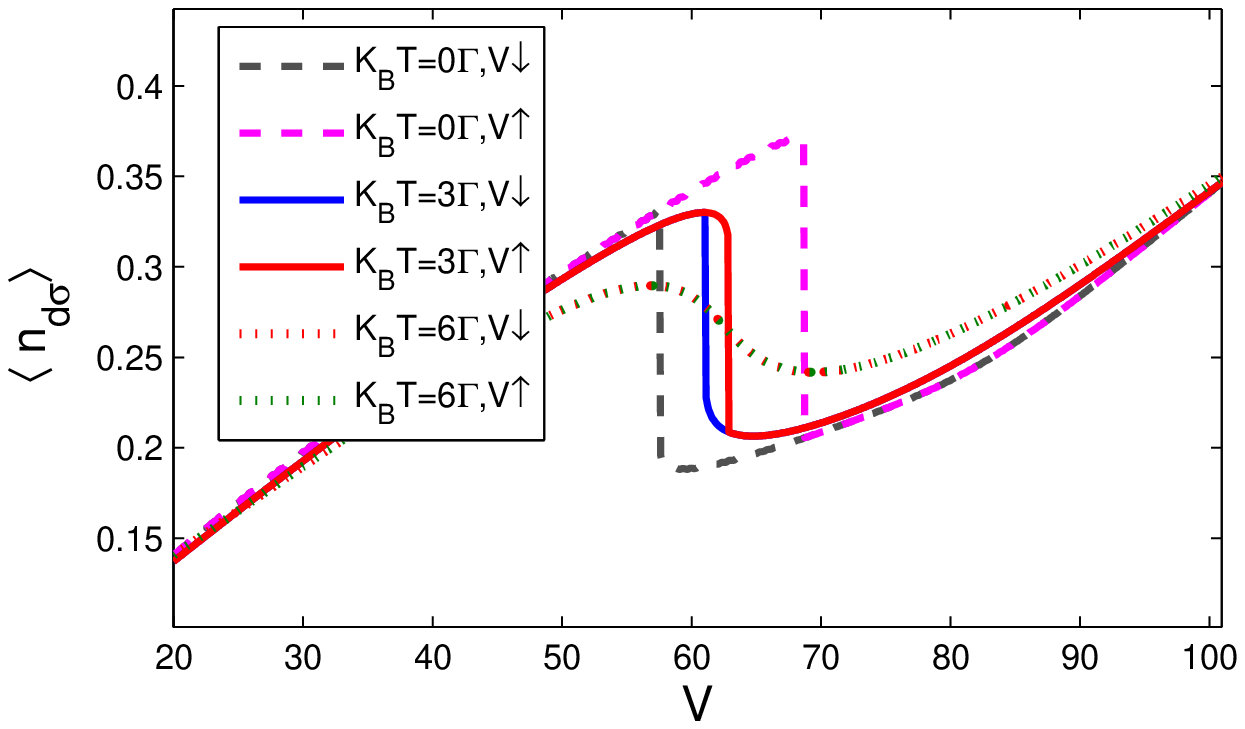}}\\
\caption{\protect\small{(Color online)(a) The average $b$-orbital occupation number $\lan n_{b\sa}\ran$ (b) The average $d$-orbital occupation number $\lan n_{d\sa}\ran$ as a function of bias $V$ (in units of $\Gamma$) for temperatures $k_BT$ equal to $0, ~3 \Gamma, ~6 \Gamma$ for a QDA of size $N=100$, with $R^d_{\rm asym}=9$ and $t_d=15 \Gamma$. The other parameters are the same as in Fig.~\ref{fig:diffALR}.  }}\label{fig:diffT}
\end{figure}

\subsection{Effect of inter $b$-orbital Coulomb interaction $\widetilde U_{xbb}$ {and of screening}\label{sec:Uxdd}}
{So far the inter $b$-orbital Coulomb interaction and the screening effect are neglected. To illustrate these two effects, we take intradot Coulomb energies $U_{bb}=U,~U_{db}=0.6U,~U_{dd}=0.5U$ and interdot Coulomb energies $U_{xdb}=U_{db}/4,~U_{xdd}=U_{dd}/4$. We first study the role of inter $b$-orbital Coulomb interaction over bistability. Next, we take into account the screening effect. \

Let us first consider the inter $b$-orbital Coulomb interaction and ignore the screening effect by setting $L_{jl}\rightarrow \infty$. In such case, we find $\widetilde U_{db}=1.31U$, $\widetilde U_{dd}=1.1U$, and $\widetilde U_{xbb}=U_{xbb}$ for QDA size $N=100$.} As shown in Fig.~\ref{fig:Uxbb} for the $b$- and $d$-orbital occupation numbers, inclusion of the Coulomb interaction between two neighboring $b$ orbitals will decrease bistability. This suppression is reflected on the sweep-down $I-V$ curve, where the population switching point gradually shifts to the right as the interdot Coulomb repulsion strength $U_{xbb}$ increases. The switching point for the sweep-up $I-V$ curve remains almost fixed.\

This phenomenon can be understood from the interplay of probability factors $\bar p_i$'s in Eq.~(\ref{eq:occpNb}) (In the considered bias range, $c_b$ is almost zero, and thus irrelevant): As we sweep upward the bias voltage, the $b$ orbital stays uncharged, i.e., $N_{b\sa}\approx 0$. This leads to $a_b\approx 1,~ b_b\approx 0$, and $c_b\approx 0$, which leaves only the probability factor $\bar p_1\approx 1$, and the rest almost zero. This makes Eq.~(\ref{eq:occpNb}) and consequently the population switching point almost unaffected by $\widetilde U_{xbb}$. However, as we begin to sweep downward the bias voltage, the fact that $N_{b\sa}\approx 1/3$ {at high bias} makes the probability factors $\bar p_1,~\bar p_2$ and $\bar p_3$ nonzero. The interplay of these three configurations (channels) leads to gradual decrease of $N_{b\sa}$, instead of a sharp fall. {Compared with the bistable range for $U_{xbb}=0$, it drops already by $85\%$ to $4.6\Gamma$ at $U_{xbb}=11.67\Gamma(U/6)$}, showing that it is sensitive to the localized nature of the $b$ orbitals. \

{To study the screening effect, we show how bistability is affected, through the suppression of interdot Coulomb interactions, by the values of screening length. From Eq.~(\ref{eq:v_dd}), we find that the effective screened Coulomb energies $\widetilde U_{db}$ and $\widetilde U_{dd}$ reduce only to $0.64U$ and $0.56U$, respectively if we take the screening lengths $L_{db}=2R_1/3$ and $L_{dd}=R_1$, where $R_1$ is the  nearest-neighbor distance. As discussed in Sec.\ref{subsec:size} and Sec.\ref{sec:CoulombU}, we expect that the reduction of $\widetilde U_{db}$ and $\widetilde U_{dd}$ will suppress bistability. However, as the impedimental inter $b$-orbital Coulomb interaction can also be screened to a much smaller value, we expect enhancement of bistability. For a shorter screening length for the localized $b$ orbital, namely $L_{bb}=R_1/2$, the screened inter $b$-orbital Coulomb energy $\widetilde U_{xbb}$ now reduces to $0.14U_{xbb}$. Thus, as shown in the dashed curves of Fig.~(\ref{fig:Uxbb}), we still find a sizable bistable range $4.5\Gamma$ even for small values of $\widetilde U_{db}$ and $\widetilde U_{dd}$.
}

\subsection{Finite temperature effect\label{sec:FiniteT}}
For practical applications, it is important to understand the temperature effect on bistability of this system. Figure \ref{fig:diffT} shows how bistability is destroyed when temperature increases. To explain the role of temperature, again we resort to the mechanism of population switching between the $b$ and $d$ orbital states. At zero temperature, when we increase the bias, the $d$ states get first populated soon as their effective energy level meets the bias-dependent chemical potential, and then are able to push by repulsive Coulomb interaction the $b$ state to higher energy. However, at finite temperatures larger than the tunnelling rates, the broadening effect due to finite tunnelling rates is taken over by the temperature which smears the characteristics of the spectral density.
When the temperature equals the effective $b$-state tunnelling rate $\tilde{\Gamma}_b$, there is no priority for which of the $d$ and $b$ states to be populated. The population switching as well as bistability will therefore disappear. In Fig.~\ref{fig:diffT}, the rate $\tilde{\Gamma}_b$ is equal to roughly $6.6\Gamma$ and $8.6\Gamma$ before and after the switching point ($V=68.6\Gamma$) at zero temperature. Thus to ensure bistability, the temperature must be much smaller than $\tilde{\Gamma}_b$.

\begin{figure}[t]
 \centering
 \subfigure[] { \label{fig:GaX2Gaz02N100_Z}
\includegraphics[scale=0.61]{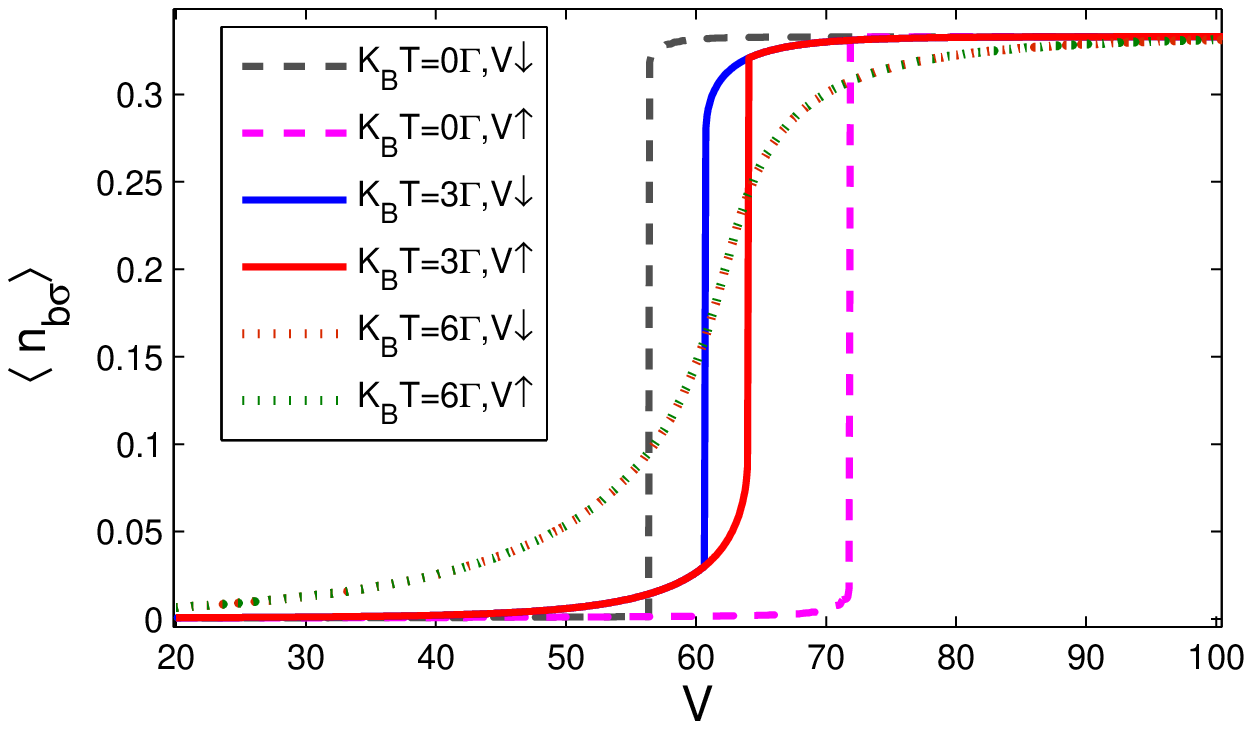} }\
\subfigure[] { \label{fig:GaX2Gaz02N100_X}
\includegraphics[scale=0.61]{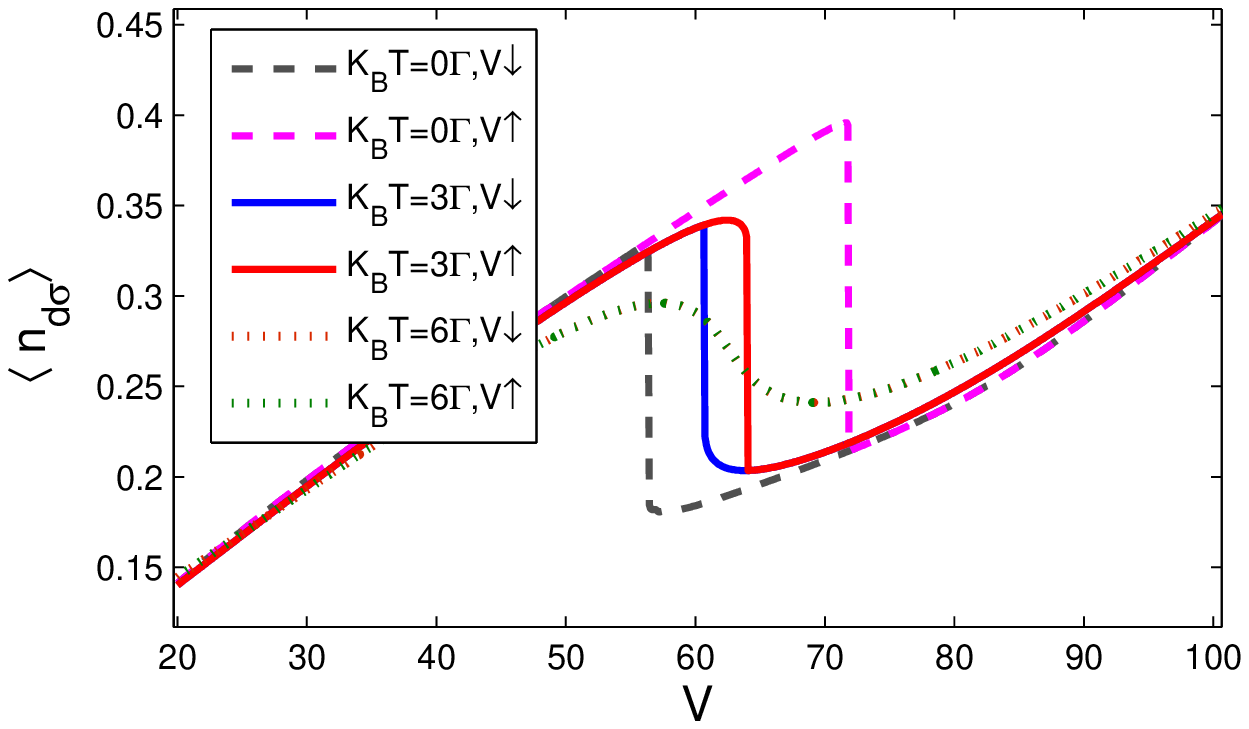}}\\
\caption{\protect\small{(Color online)(a) The average $b$-orbital occupation number $\lan n_{b\sa}\ran$ (b) The average $d$-orbital occupation number $\lan n_{d\sa}\ran$ as a function of bias $V$ (in units of $\Gamma$) for temperatures $k_BT$ equal to $0, ~3 \Gamma, ~6 \Gamma$ for a QDA of size $N=100$. In contrast with all the other figures, here the opposite limiting situation is considered, namely, we take $\Gamma_b=0.2\Gamma$, $\Gamma_d=2\Gamma$. The other parameters are the same as in Fig.~\ref{fig:diffT}.  }}\label{fig:GaX2Gaz02}
\end{figure}

\subsection{The other limiting case: $t_{\alpha b\sa}\ll t_{\alpha d\sa}$\label{sec:tblltd}}
So far bistability is discussed by using the first case in which $t_{\alpha d\sa}\ll t_{\alpha b\sa}$. As for the opposite limiting case ( $t_{\alpha b\sa}\ll t_{\alpha d\sa}$), we choose to show in Fig.~\ref{fig:GaX2Gaz02} the temperature effect on bistability, to be compared with the corresponding results in the first case as shown in Fig.~\ref{fig:diffT}.

So far as bistability is concerned, there is no qualitative difference in these two opposite limiting cases. Quantitatively, however, the case in which $t_{\alpha b\sa}\ll t_{\alpha d\sa}$ shows a larger bistable effect. At zero temperature, its bistable range in Fig.~\ref{fig:GaX2Gaz02} measures $15.3\Gamma$, larger than $11.0\Gamma$ in Fig.~\ref{fig:diffT}. This quantitative advantage goes on to finite temperatures. However, at temperatures larger than $6\Gamma$, bistability in both cases disappears. Generally, it shows that the second limiting case is more resistant to temperature, albeit within a characteristic temperature limit.

\section{Conclusions}
In this paper, a nonequilibrium Green function method\cite{KuoChang07,Chang08} is applied to study the tunnelling current through a ring of coupled quantum dots (a 1D QDA system). This method is able to treat adequately the electronic correlation in nanostructures with multiple energy levels. Bistability in the tunnelling current is found if the number of dots is large enough {and the temperature much smaller than the effective tunneling rates through the localized states}. The effects due to size scaling, Coulomb energies, {screening effect}, tight-binding interaction for the delocalized states on bistability were also investigated. Our results show that in order for bistability to occur, we need to have a) a large enough Coulomb energy $U$, b) small interdot Coulomb interaction ($\widetilde U_{xbb}$) between the localized states, and c) the tight-binding interaction between the delocalized states ($t_d$) to be around $U/4$ for best performance. {Furthermore, we showed the mixed role of the screening effect: on the one hand, it decreases bistability through reduction of the effective Coulomb interactions $\widetilde U_{db}$ and $\widetilde U_{dd}$; on the other hand, it enhances bistability through reduction of the inter $b$-orbital Coulomb interaction $\widetilde U_{xbb}$.} Most importantly, we show that bistability in the tunnelling current occurs when $N\gtrsim 50$ and becomes sizable around $N\sim 100$ in a QDA system. This size estimation provides an important information for fabrication of QDAs aimed at memory devices. Based on our analysis for a QDA of size $N=100$, in order to use the 1D QDA as a memory device at room temperature, the characteristic energy ($\Gamma$) used must be around 10 meV and a Coulomb charging energy around 700 meV, which can only be achieved in very small quantum dots (with size around 1nm). The requirement should become less stringent for 1D QDAs of larger size, and for 2D QDAs as demonstrated in the previous study\cite{KuoChang09} based on mean-field theory.

\section*{Acknowledgment}
This work was supported in part by the Academia Sinica Nano Program and National Science Council of
Taiwan under grant Nos. NSC 98-2112-M-001-022-MY3 and NSC 99-2112-M-008-018-MY2. We would like to thank Li-Chuan Tang for his assistance in numeric coding.

\renewcommand{\thesection}{\mbox{Appendix}} 
\setcounter{section}{0}

\renewcommand{\theequation}{\mbox{A.\arabic{equation}}} 
\setcounter{equation}{0} 

\section{Lesser Green functions for a two-level quantum dot\label{app_lsGF}}

In this appendix, we shall derive in detail the lesser Green functions (\ref{eq:lessGF},\ref{eq:lessGF1}) for the case of a two-level nanostructure using the same approximation mentioned in Sec.\ref{sec:GFs}. Consider two levels denoted by ${\ell}$ and $j \; (j\ne {\ell})$. Using the nonequilibrium equation of motion method\cite{Niu99}, the equation of
motion for the lesser Green function $G^{<}_{\ell,\sigma}(\va)$ reads
\bea
\mu_l G^<_{\ell,\sigma}(\va)&=&\Sigma^<_{\ell\sa}(\va)G^a_{\ell,\sigma}(\va)+U_\ell G^<_{\ell,\ell}(\va)
\nn\\
&&+ U_{\ell,j}[
G^<_{\ell,j,\bsa}(\va)+G^<_{\ell,j,\sigma}(\va)],\label{app:lsg1}
\eea
where $\mu_l \equiv
\va-\va_{\ell\sa}+i\Gamma_{\ell}/2$. The lesser self-energy is given by $\Sigma_{\ell\sa}^<(\va)=i\sum_{\alpha=L,R}\Gamma_{\alpha\ell\sa} f_\alpha(\va)$. $G^a_{\ell,\sigma}(\va)$ denotes the advanced Green function. $G^<_{\ell,\ell}(\va)=\llan
n_{\ell,\bsa}d_{\ell,\sigma},d^{\dagger}_{\ell,\sigma}\rran^<$,
$G^<_{\ell,j,\bsa}(\va)=\llan
n_{j,\bsa}d_{\ell,\sigma},d^{\dagger}_{\ell,\sigma}\rran^<$ and
$G^<_{\ell,j,\sigma}(\va)=\llan
n_{j,\sigma}d_{\ell,\sigma},d^{\dagger}_{\ell,\sigma}\rran^<$ are two-particle Green functions coupled to
$G^<_{\ell,\sigma}$ via the intralevel and interlevel Coulomb
interactions.

To solve Eq.~(\ref{app:lsg1}), we derive the equations of motion for
the two-particle Green functions. They satisfy for $\ell\not=j$
\bea
(\mu_l -U_\ell) G^<_{\ell,\ell}(\va)
&=&\Sigma^<_{\ell\sa}(\va)G^a_{\ell,\ell}(\va)+U_{\ell,j}[G^<_{\ell,\ell,j,\bsa}(\va)\nn\\
&&+G^<_{\ell,\ell,j,\sigma}(\va)]\label{app:lsg21},\\
(\mu_l -U_{\ell,j}) G^{<}_{\ell,j,\bsa}(\va)&=&\Sigma^<_{\ell\sa}(\va)G^{a}_{\ell,j,\bsa}(\va)
+U_{\ell} G^<_{\ell,\ell,j,\bsa}(\va)\nn\\
&&+U_{\ell,j}
G^<_{\ell,j,j}(\va),\label{app:lsg22}\\
(\mu_l -U_{\ell,j}) G^<_{\ell,j,\sigma}(\va)&=&\Sigma^<_{\ell\sa}(\va) G^a_{\ell,j,\sigma}(\va)
+U_{\ell}
G^<_{\ell,\ell,j,\sigma}(\va)\nn\\
&&+U_{\ell,j}G^<_{\ell,j,j}(\va).\label{app:lsg23}
\eea
Two-particle lesser Green functions are now coupled to the following
three-particle Green functions: $G^<_{\ell,\ell,j,\bsa}(\va)=\llan{n_{\ell,\bsa}
n_{j,\bsa} d_{\ell,\sigma},d^{\dagger}_{\ell,\sigma}}\rran^<$,
$G^<_{\ell,\ell,j,\sigma}(\va)=\llan{n_{\ell,\bsa}
n_{j,\sigma} d_{\ell,\sigma},d^{\dagger}_{\ell,\sigma}}\rran^<$, and
$G^<_{\ell,j,j}(\va)=\llan{n_{j,\bsa}n_{j,\sigma}
d_{\ell,\sigma},d^{\dagger}_{\ell,\sigma}}\rran^<$.
These three-particle Green functions will again couple via
their equations of motion with the four-particle Green functions, where the hierarchy
terminates. Three-particle Green functions are given by
\bea
(\mu_l-U_\ell-U_{\ell,j})G^<_{\ell,\ell,j,\bsa}(\va)
&=&\Sigma^<_{\ell\sa}(\va)G^a_{\ell,\ell,j,\bsa}(\va) \nn\\
&&+ U_{\ell,j}G^<_4(\va) \label{app:lsg31},\\
(\mu_l-U_\ell-U_{\ell,j})G^<_{\ell,\ell,j,\sigma}(\va)
&=&\Sigma^<_{\ell\sa}(\va)G^a_{\ell,\ell,j,\sigma}(\va)\nn\\
&&+ U_{\ell,j}G^<_4(\va) , \label{app:lsg32}\\
(\mu_l-2U_{\ell,j})G^<_{\ell,j,j}(\va) &=&\Sigma^<_{\ell\sa}(\va)G^a_{\ell,j,j}(\va)\nn\\
&&+ U_{\ell}G^<_4(\va), \label{app:lsg33}
\eea
where $G^<_4(\va)= \llan n_{\ell,\bsa} n_{j,\bsa}
n_{j,\sigma}d_{\ell,\sigma},d^{\dagger}_{\ell,\sigma} \rran^< $ is
the four-particle Green's function whose equation of motion reads
\bea
(\mu_l-U_{\ell}-2U_{\ell,j})G^<_4(\va) &=&\Sigma^<_{\ell\sa}(\va) G^a_4(\va) .
\eea
To solve the closed set of lesser Green functions requires solutions of the advanced Green functions $G^a$'s.
Their complex conjugates, namely the retarded Green functions, can be found in Appendix. A of Ref.~\onlinecite{Chang08}.
With $G^a_4(\va)$ known, we immediately see
\bea
G^<_4(\va) &=&\Sigma^<_{\ell\sa}(\va) \frac{N_{\ell,\bsa} c_j}{|\mu_l-U_{\ell}-2U_{\ell,j}|^2}.
\eea
Inserting the above equation and solutions of $G^a_{\ell,\ell,j,\bsa}(\va)$, $G^a_{\ell,\ell,j,\sigma}(\va)$ and $G^a_{\ell,j,j}(\va)$ into Eqs.~(\ref{app:lsg31},\ref{app:lsg32},\ref{app:lsg33}) yields three-particle Green functions as
\bea
G^<_{\ell,\ell,j,\bsa}(\va)& =& \Sigma^<_{\ell\sa}(\va)N_{\ell,\bsa} \Big[
\frac{N_{j,\bsa}-c_j}{|\mu_l-U_{\ell}-U_{\ell,j}|^2} \nn\\
&&+\frac{c_j} {|\mu_l-U_{\ell}-2U_{\ell,j}|^2} \Big]\label{app:lsg31_1},
\eea
\begin{eqnarray}
G^<_{\ell,j,j}(\va)=\Sigma^<_{\ell\sa}(\va) c_j \Big[
\frac{1-N_{\ell,\bsa}}{|\mu_l-2U_{\ell,j}|^2}\nn\\
+\frac{N_{\ell,\bsa}}{|\mu_l-U_{\ell}-2U_{\ell,j}|^2}\Big],\label{app:lsg32_1}
\end{eqnarray}
and
\bea
G^<_{\ell,\ell,j,\sigma}(\va)+G^<_{\ell,\ell,j,\bsa}(\va)=\Sigma^<_{\ell\sa}(\va)N_{\ell,\bsa} \Big[ \frac{b_j}{|\mu_l-U_{\ell}-U_{\ell,j}|^2} \nn
\eea
\bea
+\frac{2c_j} {|\mu_l-U_{\ell}-2U_{\ell,j}|^2} \Big] ,\label{app:lsg33_1}
\eea
where $b_j, ~c_j$ as well as $a_j$ are defined in the main text.
Inserting Eq.~(\ref{app:lsg33_1}) and the solution of $G^a_{\ell,\ell}(\va)$ into Eq.~(\ref{app:lsg21}) yields
\begin{eqnarray}
G^{<}_{\ell,\ell}(\va)&=&  \Sigma^<_{\ell\sa}(\va)N_{\ell,\bsa}\Big[ \frac {a_j}
{|\mu_l-U_{\ell}|^2}+ \frac{b_j }{|\mu_l-U_{\ell}-U_{\ell,j}|^2}  \nonumber \\
&& +
\frac{c_j }{|\mu_l-U_{\ell}-2U_{\ell,j}|^2} \Big],\label{app:lsg21_1}
\end{eqnarray}
while inserting Eqs.~(\ref{app:lsg32_1}), (\ref{app:lsg33_1}) and solutions of $G^{a}_{\ell,j,\bsa}(\va)$, $G^a_{\ell,j,\sigma}(\va)$
into Eqs.~(\ref{app:lsg22}) and (\ref{app:lsg23}), we obtain
\begin{eqnarray}
G^{<}_{\ell,j,\bsa}(\va)+G^{<}_{\ell,j,\sigma}(\va)=
 \Sigma^<_{\ell\sa}(\va)\Big\{b_j \Big[ \frac{1-N_{\ell,\bsa}}
{|\mu_l-U_{\ell,j}|^2} \nonumber \\
+ \frac{N_{\ell,\bsa}}{|\mu_l-U_{\ell}-U_{\ell,j}|^2}\Big] + 2 c_j \Big[\frac{1-N_{\ell,\bsa}} {|\mu_l-2U_{\ell,j}|^2} \nn\\
+\frac{N_{\ell,\bsa}}{|\mu_l-U_{\ell}-2U_{\ell,j}|^2} \Big]\Big\}.\label{app:lsg23_1}
\end{eqnarray}

Combining Eqs.~(\ref{app:lsg21_1}), (\ref{app:lsg23_1}), (\ref{app:lsg1}) and the solution of $G^a_{\ell,\sigma}(\va)$, we obtain finally
 \bea
 G^<_{\ell,\sigma}(\va)=\Sigma^<_{\ell\sa}(\va)\sum_{m=1}^3 p_m\Big[\frac
{(1-N_{\ell,\bsa})} {|\mu_l-\Pi_m|^2} \nn\\
+ \frac {N_{\ell,\bsa}}
{|\mu_l-U_{\ell}-\Pi_m|^2}\Big],
\eea
where $\Pi_1=0, \Pi_2=U_{\ell,j}$, and
$\Pi_2=2U_{\ell,j}$; $p_1=a_j$, $p_2=b_j$, and $p_3=c_j$.

From the expressions for all these lesser Green functions, we see that they all take the form of Eqs.~(\ref{eq:lessGF},\ref{eq:lessGF1}), namely,
\bea
G^<&=&-2\frac{\Sigma^<_{\ell\sa}(\va)}{\Gamma_\ell}{\rm Im} G^r\nn\\
&=&-2i\frac{\Gamma_{L\ell\sa} f_L(\va)+\Gamma_{R\ell\sa} f_R(\va)}{\Gamma_\ell}{\rm Im} G^r.
\eea

\end{document}